\documentclass[journal]{IEEEtran}

\usepackage{graphicx}
\usepackage{color}
\usepackage{amsfonts}
\usepackage{float}
\usepackage{amsmath}
\usepackage{amsthm}
\usepackage{epstopdf}
\usepackage{changepage}
\usepackage{latexsym}
\usepackage{tikz}
\usetikzlibrary{arrows}
\usetikzlibrary{automata}
\usetikzlibrary{er}
\usetikzlibrary{positioning}
\usetikzlibrary{shapes,arrows}
\usetikzlibrary{calc}
\usepackage{relsize}
\usepackage{scalefnt}
\usepackage[margin=10pt,font=footnotesize,justification=centering]{subcaption} 
\usepackage{caption}
\usepackage[margin=10pt,font=footnotesize,justification=centering,labelsep=period]{caption}
\usepackage{relsize}

\tikzset{fontscale/.style = {font=\relsize{#1}}}

\newcommand{\tabincell}[2]{\begin{tabular}{@{}#1@{}}#2\end{tabular}}

\begin{document}

\title{An Interaction Model for Simulation and \\ Mitigation of Cascading Failures}

\author{Junjian~Qi,~\IEEEmembership{Member,~IEEE,}
        Kai~Sun,~\IEEEmembership{Senior Member,~IEEE,}
        and Shengwei Mei,~\IEEEmembership{Senior Member,~IEEE}%
        \thanks{J.~Qi and K.~Sun are with Dept. of Electrical Engineering \& Computer Science, University of Tennessee, Knoxville, TN 37996 (e-mails: junjian.qi.2012@ieee.org and kaisun@utk.edu). 
S.~Mei is with the State Key Laboratory of Power Systems, Dept. of Electrical Engineering, Tsinghua University, Beijing China 100084 (e-mail: meishengwei@mail.tsinghua.edu.cn).
This material is based upon work supported partially by the University of Tennessee in Knoxville, the CURENT Engineering Research Center, and NSFC grant-51321005.
}}

\markboth{preprint of  DOI 10.1109/TPWRS.2014.2337284, IEEE Transactions on Power Systems.}{stuff}\maketitle

\begin{abstract}
In this paper the interactions between component failures are quantified and the interaction matrix and interaction network are obtained. 
The quantified interactions can capture the general propagation patterns of the cascades from utilities or simulation, thus helping to better understand how cascading failures propagate and to identify key links and key components that are crucial for cascading failure propagation. By utilizing these interactions a high-level probabilistic model called interaction model is proposed 
to study the influence of interactions on cascading failure risk and to support online decision-making.
It is much more time efficient to first quantify the interactions between component failures 
with fewer original cascades from a more detailed cascading failure model and then perform the interaction model simulation than it is to directly simulate a large number of cascades with a more detailed model.
Interaction-based mitigation measures are suggested to mitigate cascading failure risk by weakening key links, which can be achieved in real systems by wide area protection such as blocking of some specific protective relays.
The proposed interaction quantifying method and interaction model are validated with line outage data generated by 
the AC OPA cascading simulations on the IEEE 118-bus system.
\end{abstract}


\section{Introduction}

\IEEEPARstart{C}{ascading} blackouts are complicated sequences of dependent outages which could 
bring about tremendous economic and social losses [\ref{nerc}]--[\ref{TF}].
Large rare cascading blackouts have substantial risk and pose great challenges in simulation, analysis, and mitigation.
It is important to study the mechanisms of cascading failures 
so that the risk of large-scale blackouts may be better quantified and mitigated.

In order to study cascading failures, several models have been proposed, such as
CASCADE model [\ref{cascade}], branching process model [\ref{bp10}]--[\ref{bp12}], hidden failure model [\ref{Phadke}], [\ref{chen}], 
OPA\footnote{\textbf{OPA} stands for \textbf{O}ak Ridge National Laboratory, \textbf{P}ower Systems Engineering Research Center at the University of Wisconsin, University of \textbf{A}laska to indicate the institutions collaborating to devise the simulation.} 
model [\ref{opa1}]--[\ref{opa4}], improved OPA model [\ref{improved}], AC OPA model [\ref{AC OPA}], [\ref{AC OPA1}], OPA with slow process [\ref{slow}], Manchester model [\ref{Rios}], [\ref{Kirschen}], stochastic model [\ref{Anghel}], dynamic PRA model [\ref{pra}], and influence model \cite{influence}.

The OPA model [\ref{opa1}]--[\ref{opa4}] simulates the patterns of cascading blackouts of a power
system under the complex dynamics of a growing demand and the engineering responses to failure.
Initial line outages are generated randomly by assuming that each line can fail independently.
Whenever a line fails, the generation and load is redispatched with DC load flow model and linear programming. 
If any lines were overloaded during the optimization, then these lines are those that are likely to have experienced high stress, and each of these lines fails independently. The process of redispatch and testing for line outages is iterated until there are no more outages.

The branching process model [\ref{bp10}]--[\ref{bp12}] can statistically describe the statistical or simulated cascades and provide higher-level statistical information about cascading failures by tracking the numbers of lines outaged and the amounts of load shed. But it does not retain information about the network topology or load flow and also does not attempt to specify how cascades propagate in the system in detail, such as which, where, or why lines outage. The most recent study on the line interaction graph \cite{line graph} 
initiates a novel analysis method for cascading failures by considering the interactions of transmission lines 
and tries to understand cascading failures with models amenable to analysis while keeping the basic physics of power systems.

From the perspective of complex systems the system-level failures are not caused by any specific event but by the property that the components in the system are tightly coupled and interdependent \cite{normalAccident}.
Thus explicitly studying the interactions between components can help understand the mechanisms of cascading failures, identify the key factors for their propagation, and further propose effective mitigation measures.

In this paper we quantify the interactions between component failures by following the line graph approach in \cite{line graph}. These interactions can capture the general patterns of the propagation of cascading failures in a system and help better understand why and how cascading failures occur and propagate. Key links between component failures which play important roles in the propagation of cascading failures can also be identified and further be used for determining wide area protection schemes [\ref{begovic}]--[\ref{sun2}], such as relay blocking under the condition of some specific line tripping, which can secure time to perform remedial controls by a defense system during cascaded events [\ref{CCLiu}].
A cascading failure model called interaction model is further proposed based on these interactions to speed up simulation and to study how component interactions influence cascading failure risks.
It is much more time efficient to first quantify the interactions between the component failures 
with fewer original cascades from more detailed cascading failure model, such as AC OPA [\ref{AC OPA}], [\ref{AC OPA1}], and then perform the interaction model simulation than it is to directly simulate a large number of cascades with a more detailed model.

Besides, topological properties such as small-world [\ref{small world}] 
and scale-free [\ref{scale free}] behaviors have been found in complex networks.
But it can be misleading to evaluate the vulnerability of power systems only with topological metrics [\ref{topology}].
In this paper we discuss the property of a directed weighted interaction network 
generated with simulated cascades from a more detailed cascading failure model
which considers the physics of the system such as power flow and re-dispatching
rather than directly exploring the property of the network from the topology of the physical system.

The rest of this paper is organized as follows. 
Section \ref{interaction} explains how the interactions between component failures can be quantified.
Section \ref{keyLink} discusses the identification of the key links and key components based on the obtained interactions.
Section \ref{model} proposes an interaction model by using the quantified interactions and also discusses methods for validating it.
Section \ref{num needed} proposes methods to determine how many cascades should be simulated and how many cascades should be utilized to quantify the interaction between component failures.
Section \ref{miti} discusses the mitigation measures by weakening key links and the potential application in real systems as a wide area protection scheme.
Section \ref{simulation} tests the proposed interaction quantifying method and the interaction model with line outage data generated by AC OPA simulations 
on IEEE 118-bus system. Finally the conclusion is drawn in Section \ref{conclusion}.

\section{Quantifying the Interactions between Component Failures} \label{interaction}

In this section the cascades that record cascading failure sequences
are used to quantify the interactions between component failures.

For power systems the transmission lines or transformers can be chosen as components 
and the cascades can come from either statistical utility line outage data or simulations 
from more detailed cascading failure models.
The statistical data can be grouped into different cascades and 
then into different generations within each cascade based on outages' timing [\ref{bp12}].
The simulation data can be generated from OPA model or its variants [\ref{opa1}]--[\ref{slow}], 
which naturally produce line outages in generations or stages; each iteration of the ``main loop" of the simulation produces another generation [\ref{bp10}], [\ref{bp13}].
AC OPA model [\ref{AC OPA}], [\ref{AC OPA1}] will be employed in the case studies of 
this paper to produce cascades for qualifying component interactions.
But the data can also be generated from other cascading simulation tools as long as
they can be grouped into cascades and generations.

The cascades used for quantifying the interactions between component failures are called original cascades
in order to distinguish the simulated cascades from the proposed model in this paper.
$M$ original cascades can be arranged as
\begin{table}[H]
\renewcommand{\arraystretch}{2.0}
\label{data}
\centering
\begin{tabular}{ccccc}
& generation\,0 & generation\,1 & generation\,2 & $\cdots$ \\
\textrm{cascade 1} & $F^{(1)}_{0}$ & $F^{(1)}_{1}$ & $F^{(1)}_{2}$ & $\cdots$ \\
\textrm{cascade 2} & $F^{(2)}_{0}$ & $F^{(2)}_{1}$ & $F^{(2)}_{2}$ & $\cdots$ \\
$\vdots$ & $\vdots$ & $\vdots$ & $\vdots$ & $\vdots$ \\
\textrm{cascade} $M$ & $F^{(M)}_{0}$ & $F^{(M)}_{1}$ & $F^{(M)}_{2}$ & $\cdots$
\end{tabular}
\end{table}
\noindent
where $F^{(m)}_{g}$ is the set of failed components 
produced in generation $g$ of cascade $m$.
Each cascade eventually terminates with a finite number of generations 
when the number of failed components in a generation becomes zero.
The shortest cascades stop in generation 1 by having no outages in generation 1 and higher generations,
but some of the cascades will continue for several or occasionally many generations before terminating.

After obtaining $M$ original cascades we can quantify the interactions between component failures
based on all or part of them. 
Assume $M_u \le M$ original cascades are utilized to quantify the interactions. 
Since at first we do not have enough information to
determine which components in two consecutive generations have interactions, 
we assume that there are interactions between any failed component 
in last generation and that in this generation to guarantee that no interactions will be ignored.
Thus for a system with $n$ components, a matrix $\boldsymbol{A}\in \mathbb{Z}^{n\times n}$ 
can be constructed, whose entry $a_{ij}$ is the number of times that
component $i$ fails in one generation before the failure of component $j$
among all original cascades.
Since $\boldsymbol{A}$ is obtained by using all $M_u$ cascades it does not depend on the order that the cascades are processed.

The assumption based on which $\boldsymbol{A}$ is obtained actually exaggerates the interactions between component failures since it is not convincing to assert one component interacts with another one only because it fails in its last generation. Therefore, for each failed component in generation 1 and the following generations 
the failed component that most probably causes it should be determined.

Specifically, for any two consecutive generations $k$ and $k+1$ of any cascade $m$, 
the failure of component $j$ in generation $k+1$ is considered to be caused 
by a set of failed components in generation $k$, which can be described as
\begin{equation}
\{i_c|i_c\in F_k^{(m)}\; \textrm{and} \; a_{i_c j}=\max_{i\in F_k^{(m)}} a_{ij}\}.
\end{equation}
Note that it is possible that two or more components in generation $k$ are
considered as the cause of the failure of component $j$. 
When $M_u$ is not large enough this will be more possible because in this case 
no component has much greater $a_{ij}$ than the others. In the extreme case for which all $a_{ij}$ for $i\in F_k^{(m)}$ are the same it will become impossible to determine which component more possibly causes the failure of component $j$ and thus all components will be considered as the cause.
Then the overestimation of the interaction by $\boldsymbol{A}$ cannot be well corrected, which will lead to 
the overestimation of the propagation of cascading failures.
This will be discussed further in section \ref{cascade needed results}.

\begin{figure}[!t]
\captionsetup{justification=raggedright,singlelinecheck=false}
\hspace*{-2.5cm}
\centering
\scalefont{0.9}
\begin{subfigure}[b]{0.23\textwidth}
\begin{tikzpicture}[->,>=stealth',shorten >=1pt,auto,node distance=1.8cm,on grid,semithick,
every state/.style={fill=white,draw=black,circle,scale=0.55,text=black}]
\node[state] (B) {$B$};
\node[state] (A) [left=of B] {$A$};
\node[state] (C) [right=of B] {$C$};
\node[state] (D) [below left=of B] {$D$};
\node[state] (E) [below right=of B] {$E$};
\node[anchor=west, right] at (2.5,0){\footnotesize generation $k$};
\node[anchor=west, right] at (2.5,-1.3){\footnotesize generation $k+1$};
\path (A) edge [right] node {} (D)
(A) edge [right] node {} (E)
(B) edge [left] node {} (D)
(B) edge [right] node {} (E)
(C) edge [left] node {} (D)
(C) edge [left] node {} (E);
\end{tikzpicture}
\subcaption{}
\label{fig1a}
\end{subfigure}

\vspace{0.2cm}

\hspace*{-2.4cm}
\begin{tikzpicture}
  \pgfsetarrowsstart{latex}
  \pgfsetlinewidth{1ex} 
  \pgfpathmoveto{\pgfpointorigin} 
  \pgfpathlineto{\pgfpoint{0cm}{1cm}} 
  \pgfusepath{stroke}
\end{tikzpicture}

\vspace{0.2cm}
\centering
\hspace*{-2.3cm}
\begin{subfigure}[b]{0.23\textwidth}
\begin{tikzpicture}[->,>=stealth',shorten >=1pt,auto,node distance=1.8cm,on grid,semithick,
every state/.style={fill=white,draw=black,circle,scale=0.55,text=black}]
\node[state] (B) {$B$};
\node[state] (A) [left=of B] {$A$};
\node[state] (C) [right=of B] {$C$};
\node[state] (D) [below left=of B] {$D$};
\node[state] (E) [below right=of B] {$E$};
\node[anchor=west, right] at (2.5,0){\footnotesize generation $k$};
\node[anchor=west, right] at (2.5,-1.3){\footnotesize generation $k+1$};
\path (A) edge [right] node {} (D)
(B) edge [left] node {} (D)
(C) edge [left] node {} (E);
\end{tikzpicture}
\caption{}
\label{fig1b}
\end{subfigure}
\caption{Illustration for determining the cause of component failures.}
\label{illustration}
\end{figure}
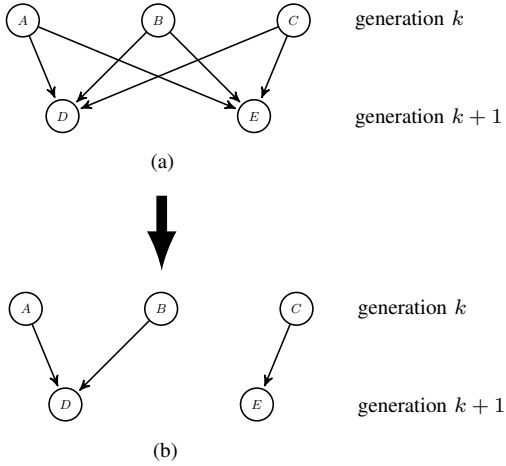

An illustration is shown in Fig. \ref{illustration}, in which we show two consecutive generations of a cascade.
If we assume that
\begin{align}
a_{AD}&=a_{BD}=\max_{i \in \{A,B,C\}}a_{iD} \\
a_{CE}&=\max_{i\in \{A,B,C\}}a_{iE}
\end{align}
we can determine the cause of component $D$ failure as component $A$ and $B$ and the cause of component $E$ failure as component $C$.

What should be emphasized here is that $\boldsymbol{A}$ is not updated when determining the most possible causes for failed components in generation 1 and the following generations. 
Therefore, the determination of the component that causes a failed component does not depend on 
the order that the original cascades are processed but is completely determined by the $\boldsymbol{A}$ matrix.

After determining the cause of any component failure in generation 1 and the following generations 
for all cascades $\boldsymbol{A}$ can be corrected to be $\boldsymbol{A}' \in \mathbb{Z}^{n\times n}$, 
whose entry $a'_{ij}$ is the number of times that the failure of component $i$ causes the failure of component $j$.

The interaction matrix $\boldsymbol{B}\in \mathbb{R}^{n\times n}$
can be calculated from $\boldsymbol{A}'$.
Its entry $b_{ij}$ is the empirical probability
that the failure of component $i$ causes the failure of component $j$, which can be given by
\begin{equation}
b_{ij}=\frac{a'_{ij}}{N_i}
\end{equation}
where $N_i$ is the number of failures of component $i$.

The $\boldsymbol{B}$ matrix determines how components interact with each other.
The nonzero elements of $\boldsymbol{B}$ are called links.
Link $l:\,i\rightarrow j$ corresponds to $\boldsymbol{B}$'s nonzero element $b_{ij}$
and starts from component $i$ and ends with component $j$.
By putting all links together a directed network $\mathcal{G}(\mathcal{C},\mathcal{L})$ called 
interaction network can be obtained. Its vertices $\mathcal{C}$ are components
and each directed link $l\in \mathcal{L}$ represents that a failure of the source vertex component
causes the failure of the destination vertex component with probability greater than 0.

\section{Identifying Key Links and Key Components} \label{keyLink}

The links can vary significantly with respect to their roles in the propagation of cascading failures.
In order to distinguish them and to further identify key links, 
an index $I_l$ is defined for each link $l:\,i\rightarrow j$
to be the expected value of the number of failures that are propagated through link $l$.
Note that the failures propagated through link $l$ can be directly triggered by the failure of component $i$
or can be triggered by the failures of components other than $i$ which finally is able to cause component $i$ to fail.

Therefore, in order to calculate $I_l$ the number of failures of its source vertex $i$, which is denoted by $N_i^s$, should be set to be $N_i$, which is the total number of failures among all the original cascades. 
$N_i$ contains not only the failures in generation 0, which serve as trigger of cascading failures, but also the failures caused by other component failures.

Similar to section \ref{interaction}, $M_u \le M$ original cascades are utilized to quantify the interactions
and further to calculate the index $I_l$.
After obtaining the interaction network we can get a directed acyclic subgraph $\mathcal{G}_j(\mathcal{C}_j,\mathcal{L}_j)$ starting with component $j$ from the interaction network $\mathcal{G}$.
The vertices represent the events of component failures and the edges represent causal relations between events.
All edges in the subgraph point in the same direction from parent to child 
due to causality affecting the future and all components are reached exactly once.

We would like to emphasize that for each link there is a unique directed acyclic subgraph which can be extracted from the whole interaction network and is comprised of all the components influenced by this link.

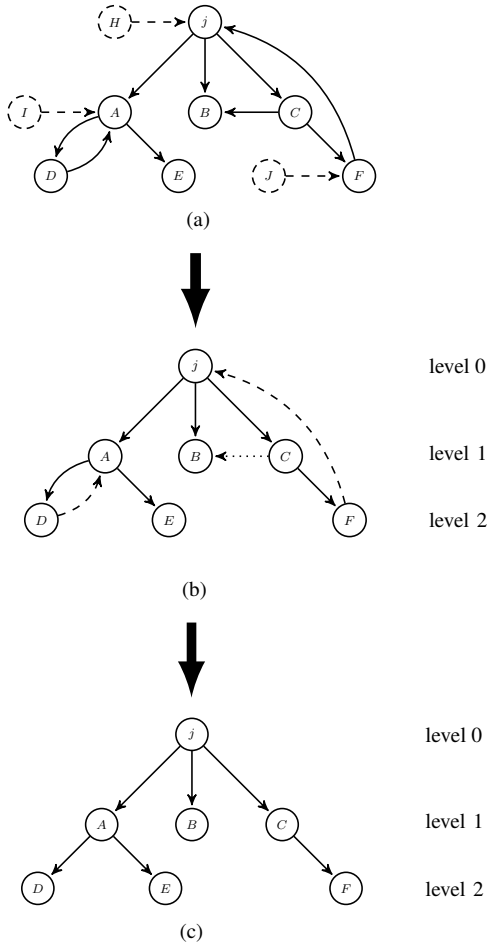
\begin{figure}[!t]
\captionsetup{justification=raggedright,singlelinecheck=false}
\hspace*{-1.55cm}
\centering
\scalefont{0.9}
\begin{subfigure}[b]{0.28\textwidth}
\begin{tikzpicture}[->,>=stealth',shorten >=1pt,auto,node distance=1.2cm,on grid,semithick,
every state/.style={fill=white,draw=black,circle,scale=0.55,text=black}]
\node[state] (j) {$j$};
\node[state] (B) [below=of j] {$B$};
\node[state] (A) [left=of B] {$A$};
\node[state] (C) [right=of B] {$C$};
\node[state] (D) [below left=of A] {$D$};
\node[state] (E) [below right=of A] {$E$};
\node[state] (F) [below right=of C] {$F$};
\node[state] (H) [densely dashed,left=of j] {$H$};
\node[state] (I) [densely dashed,left=of A] {$I$};
\node[state] (J) [densely dashed,left=of F] {$J$};
\path (j) edge [below] node {} (B)
(j) edge [left] node {} (A)
(j) edge [right] node {} (C)
(A) edge [bend right] node {} (D)
(A) edge [right] node {} (E)
(C) edge [below] node {} (F)
(C) edge [left] node {} (B)
(D) edge [bend right] node {} (A)
(F) edge [bend right] node {} (j)
(H) edge [dashed,right] node {} (j)
(I) edge [dashed,right] node {} (A)
(J) edge [dashed,right] node {} (F);
\end{tikzpicture}
\caption{}
\label{fig:a}
\end{subfigure}

\vspace{0.2cm}

\hspace*{-1.5cm}
\begin{tikzpicture}
  \pgfsetarrowsstart{latex}
  \pgfsetlinewidth{1ex} 
  \pgfpathmoveto{\pgfpointorigin} 
  \pgfpathlineto{\pgfpoint{0cm}{1cm}} 
  \pgfusepath{stroke}
\end{tikzpicture}

\vspace{0.2cm}
\hspace*{-1.6cm}
\centering
\scalefont{0.9}
\begin{subfigure}[b]{0.25\textwidth}
\begin{tikzpicture}[->,>=stealth',shorten >=1pt,auto,node distance=1.2cm,on grid,semithick,
every state/.style={fill=white,draw=black,circle,scale=0.62,text=black}]
\node[state] (j) {$j$};
\node[state] (B) [below=of j] {$B$};
\node[state] (A) [left=of B] {$A$};
\node[state] (C) [right=of B] {$C$};
\node[state] (D) [below left=of A] {$D$};
\node[state] (E) [below right=of A] {$E$};
\node[state] (F) [below right=of C] {$F$};
\node[anchor=west, right] at (3,0){level 0};
\node[anchor=west, right] at (3,-1.15){\footnotesize level 1};
\node[anchor=west, right] at (3,-2.05){\footnotesize level 2};
\path (j) edge [below] node {} (B)
(j) edge [left] node {} (A)
(j) edge [right] node {} (C)
(A) edge [bend right] node {} (D)
(A) edge [right] node {} (E)
(C) edge [below] node {} (F)
(C) edge [dotted,left] node {} (B)
(D) edge [dashed,bend right] node {} (A)
(F) edge [dashed,bend right] node {} (j);
\end{tikzpicture}
\hspace*{-1.5cm}
\caption{}
\label{fig:b}
\end{subfigure}

\vspace{0.2cm}

\hspace*{-1.6cm}
\begin{tikzpicture}
  \pgfsetarrowsstart{latex}
  \pgfsetlinewidth{1ex} 
  \pgfpathmoveto{\pgfpointorigin} 
  \pgfpathlineto{\pgfpoint{0cm}{1cm}} 
  \pgfusepath{stroke}
\end{tikzpicture}

\vspace{0.2cm}
\hspace*{-1.6cm}
\centering
\begin{subfigure}[b]{0.25\textwidth}
\begin{tikzpicture}[->,>=stealth',shorten >=1pt,auto,node distance=1.2cm,on grid,semithick,
every state/.style={fill=white,draw=black,circle,scale=0.6,text=black}]
\node[state] (j) {$j$};
\node[state] (B) [below=of j] {$B$};
\node[state] (A) [left=of B] {$A$};
\node[state] (C) [right=of B] {$C$};
\node[state] (D) [below left=of A] {$D$};
\node[state] (E) [below right=of A] {$E$};
\node[state] (F) [below right=of C] {$F$};
\node[anchor=west, right] at (3,0){level 0};
\node[anchor=west, right] at (3,-1.15){\footnotesize level 1};
\node[anchor=west, right] at (3,-2.05){\footnotesize level 2};
\path (j) edge [below] node {} (B)
(j) edge [left] node {} (A)
(j) edge [right] node {} (C)
(A) edge [left] node {} (D)
(A) edge [right] node {} (E)
(C) edge [below] node {} (F);
\end{tikzpicture}
\caption{}
\label{fig:c}
\end{subfigure}

\caption{Illustration for obtaining the directed acyclic subgraph starting with $j$.}
\label{subgraph}
\end{figure}

In Fig. \ref{subgraph} we illustrate how the directed acyclic subgraph $\mathcal{G}_j$ 
(Fig. \ref{fig:c}) can be obtained from the original subgraph (Fig. \ref{fig:a}).
Note that $i$ is not in Fig. \ref{fig:a} even if there is a link from $j$ or any other vertex
because $I_l$ is defined to indicate the failures link $l:i\rightarrow j$ can cause on the condition that 
$i$ fails.
From Fig. \ref{fig:a} to Fig. \ref{fig:b} we remove the vertices 
for which there is no path from vertex $j$ to them ($H$, $I$, and $J$ denoted by dashed circles in Fig. \ref{fig:a}) since we would like to quantify the consequences 
brought about by $j$ and the removed vertices cannot be influenced by $j$. 
The links corresponding to the removed vertices (denoted by dashed arrows in Fig. \ref{fig:a}) are also eliminated. 

In Fig. \ref{fig:b} vertex $j$ is at level 0; the vertices that $j$ points to and are not $i$ are at level 1; 
the vertices that the level 1 vertices point to and are not $i$ or any other vertex in the lower levels
are at level 2. Because of the causal relationship between the vertices in two consecutive levels
the edges from vertices at a higher level to those at a lower level ($D\rightarrow A$ and $F\rightarrow j$ denoted by dashed arrows in Fig. \ref{fig:b}) are removed. Also the edges between vertices at the same level ($C\rightarrow B$ denoted by dotted arrow in Fig. \ref{fig:b}) are neglected since these vertices are considered independent and all fail on the condition of the failure of some component at the last level. Finally we can get a directed acyclic subgraph $\mathcal{G}_j$ (Fig. \ref{fig:c}) for which there is no loop and for each vertex (component) $c \in \mathcal{C}_j,\,c\neq j$ there is exactly one vertex $c_s$ pointing to it.

The expected value of the number of failures of component $j$
given $N_i$ times of component $i$ failure is
\begin{equation}
E_j=N_i^s\,b_{ij}.
\end{equation}
For any other component $c \in \mathcal{C}_j,\,c\neq j$, the expected value of the number of failures 
given the times of its source vertex failure is
\begin{equation}
E_c=E_{c_s}\,b_{c_s j}.
\end{equation}
All the expected number of component failures in graph $\mathcal{G}_j$ are summated to be $I_l$ as
\begin{equation}
I_l=\sum_{c\in \mathcal{C}_j}E_c.
\end{equation}

$I_l$ can indicate the contribution of a link to the propagation of cascading failures.
The greater the index is, the more important the link is for cascading failure propagation.
Thus the links with large $I_l$ can be defined as key links.
Specifically, the set of key links $\mathcal{L}^{\textrm{key}}$ are those links whose weights are greater than or equal to a specified fraction of 
the largest link weight $I_l^{\textrm{max}}$, that is
\begin{equation}
\mathcal{L}^{\textrm{key}}=\{l|I_l \ge \epsilon_l I_l^{\textrm{max}}\}
\end{equation}
where $\epsilon_l$ is taken as a value that is not too close to zero to guarantee that 
the weights of key links are not much less than the largest link weight.

By taking $I_l$ as weights of the links, we can make the interaction network $\mathcal{G}(\mathcal{C},\mathcal{L})$ 
in section \ref{interaction} to be a directed weighted network. 
The vertex out-strength and in-strength of the interaction network can be defined as follows.
\begin{align}
s_i^{\textrm{out}}&=\sum_{l \in \mathcal{L}^{\textrm{out}}(i)}I_l \\
s_i^{\textrm{in}}&=\sum_{l \in \mathcal{L}^{\textrm{in}}(i)}I_l
\end{align}
where $\mathcal{L}^{\textrm{out}}(i)$ and $\mathcal{L}^{\textrm{in}}(i)$ are respectively
the sets of links starting from and ending with vertex $i$.

The out-strength and in-strength can indicate how much an component influences and is influenced by another one.
The components with large out-strength can cause great consequences and thus are crucial for the propagation of cascading failures.
Therefore, in a similar way to the key link definition, the set of key components $\mathcal{C}^{\textrm{key}}$ is defined as
\begin{equation}
\mathcal{C}^{\textrm{key}}=\{i|s_i^{\textrm{out}} \ge \epsilon_s s_i^{\textrm{out,\,max}}\}
\end{equation}
where $s_i^{\textrm{out,\,max}}$ is the largest vertex out-strength among all vertices and $\epsilon_s$ is
used to guarantee that the out-strengths of the key components are not much less than the maximum out-strength.

\section{Interaction Model} \label{model}

In this section a cascading failure model called interaction model is proposed 
by using the tripping probability of each component in generation 0 and the interactions 
between component failures, which can be obtained either from statistical utility line outage data or 
simulations generated from OPA model or its variants.
After introducing the interaction model we also discuss how it can be validated 
by comparing its simulated cascades and the original cascades.

As in section \ref{interaction}, we assume there are a total of $M$ original cascades available.
Note that we do not necessarily need to use all the $M$ cascades but only $M_u$ of them to generate the tripping probability of each component in generation 0 and the interaction matrix 
since a smaller number of cascades can capture how frequently the components fail in generation 0 and 
how the component failures interact with each other, especially when any one of the original cascades can be 
considered as an independent and identical realization of a underling process.

\subsection{Model Design}

\begin{figure}[!t]
\captionsetup{justification=raggedright,singlelinecheck=false}
\centering
\includegraphics[width=2.8in]{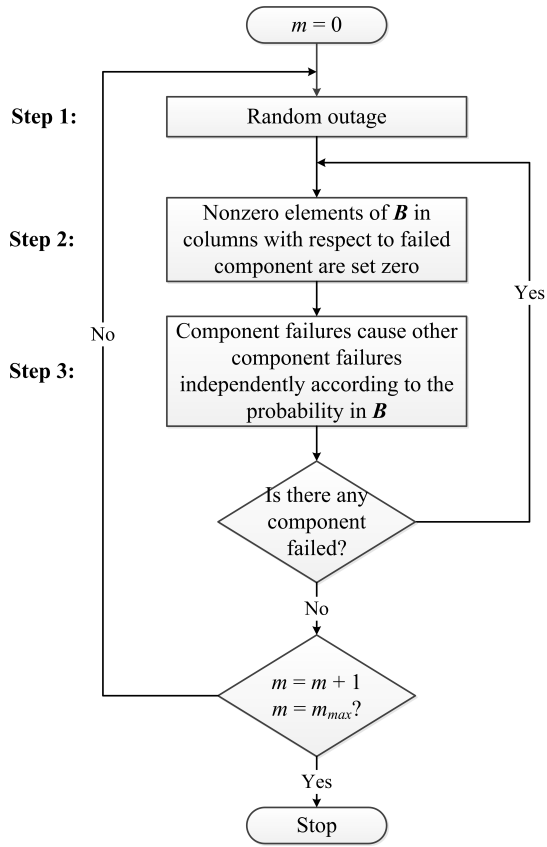}
\caption{Flow chart of the interaction model.}
\label{modelChart}
\end{figure}

It is assumed that all components are initially unfailed 
and each component fails with a small probability.
The component failures in the same generation cause other component failures independently.
The flow chart of the proposed model is shown in Fig. \ref{modelChart}, in which $m_{\textrm{max}}$ 
is the number of cascades to be simulated.

The model contains two loops and in each outer loop a cascade is simulated. 
Specifically, the model is implemented in the following 3 steps.

\vspace{0.1cm}
Step 1)	 Accidental faults of components

\vspace{0.1cm}
\begin{adjustwidth}{1.5cm}{0cm}
In the $k$th outer iteration, each component $i$ randomly fail 
with probability $\tau_i$ to simulate accidental faults and
the failed components form generation 0 (initial outages) of the simulated cascade.
The probability that a component $i$ fails as initial outages can be estimated by using 
the generation 0 component failures of the $M_u$ original cascades as
\begin{equation}
\tau_i=\frac{f_0^i}{M_u}
\end{equation}
where $f_0^i$ is the number of cascades for which component $i$ fails in generation 0.

\end{adjustwidth}

\vspace{0.1cm}
Step 2)	Corresponding columns of $\boldsymbol{B}$ are set zero

\vspace{0.1cm}
\begin{adjustwidth}{1.5cm}{0cm}
The columns of $\boldsymbol{B}$ corresponding to the component failures are set zero 
since in our model once a component fails 
it will remain that way until the end of the simulation. 
\end{adjustwidth}

\vspace{0.1cm}
Step 3)	Failed components cause other component failures

\vspace{0.1cm}
\begin{adjustwidth}{1.5cm}{0cm}
The component failures in one generation independently generate other component failures.
Specifically, if component $i$ fails in this generation
it will cause the failure of any other component $j$ with probability $b_{ij}$. 
Once it causes the failures of some components, 
these newly caused component failures will comprise the next generation; then go back to step 2. If no component failure is caused, the inner loop stops.
\end{adjustwidth}

\vspace{0.1cm}
By using the interaction model described above we can simulate as many cascades as possible (greater than $M_u$).
Although the simulated cascades are generated by utilizing the information of the initial outages 
and the interactions contained in the original cascades, they can reveal new rare events due to the high-level probabilistic property of the interaction model, thus helping recover the missing information due to using fewer original cascades.
Therefore, as long as the cascades from the interaction model are well validated it can be much more time efficient to first quantify the interactions between the component failures with fewer original cascades from a more detailed cascading failure model and then perform the interaction model simulation than it is to directly simulate a large number of cascades with a more detailed model.

We also would like to emphasize that the proposed interaction model can not only be used for 
offline study of cascading failures but can also be used for online decision-making support.
The interaction matrix can be obtained offline from statistical utility data or simulations of more detailed cascading failure models. It contains important information about the interactions between component failures. 
By utilizing this information the interaction model has the potential to predict the consequences of events. 
If something unusual happens in the system the operators can apply the interaction model to quickly find out which components or which areas of the system will most probably be affected so that a fast response can be performed to pull the system back to normal conditions and to avoid or at least reduce the economic and social losses.

\subsection{Validating the Model} \label{validate}

In order to validate the proposed interaction model, the simulated cascades from the interaction model are carefully 
compared with the original cascades by using the following four methods.

\begin{enumerate}

\item The probability distribution of total line outages of the original and simulated cascades are compared.

\item The probability distribution of total line outages of the original and simulated cascades can be estimated with the branching process and 
the average propagation (estimated offspring mean) $\hat{\lambda}$ can be compared. 
More details can be found in [\ref{bp10}], [\ref{bp13}].

\item The interactions between component failures for both the original and simulated cascades are quantified and the probability distribution of the link weights and vertex out-strength and in-strength of the interaction network are compared.

\item The spreading capacities of the links quantified from the original and simulated cascades are compared by some defined similarity indices as described below.

\end{enumerate}

For the first and third methods we only need to compare the probability distribution of the total line outages, the link weights, or the vertex out-strength and in-strength for the original and simulated cascades.
For the second method the calculation of the average propagation can be found in [\ref{bp10}], [\ref{bp13}].
Thus in the rest of this section we will only discuss the fourth method in detail.

Let $\mathcal{L}_1$, $\mathcal{L}_2$, and $\mathcal{L}_3$ be the set of links shared by the original and simulated cascades
and the links only owned by the original and simulated cascades.
Denote the index of link $l$ for the original and simulated cascades
respectively by $I_l^\textrm{ori}$ and $I_l^\textrm{sim}$.
Five similarity indices are defined as follows.
\begin{align}
S_1&=\frac{\sum\limits_{l\in (\mathcal{L}_1 \cup \mathcal{L}_3)}I_l^\textrm{sim}}
{\sum\limits_{l\in (\mathcal{L}_1 \cup \mathcal{L}_2)}I_l^\textrm{ori}} \\
S_2&=\frac{\sum\limits_{l\in \mathcal{L}_1}^{}I_l^\textrm{ori}}{\sum\limits_{l\in (\mathcal{L}_1 \cup \mathcal{L}_2)}^{}I_l^\textrm{ori}} \\
S_3&=\frac{\sum\limits_{l\in \mathcal{L}_1}^{}I_l^\textrm{sim}}{\sum\limits_{l\in (\mathcal{L}_1 \cup \mathcal{L}_3)}^{}I_l^\textrm{sim}} \\
S_4&=\frac{\sum\limits_{l\in \mathcal{L}_1}^{}I_l^\textrm{sim}}{\sum\limits_{l\in \mathcal{L}_1}^{}I_l^\textrm{ori}} \\
S_5&=\sum\limits_{l\in \mathcal{L}_1}^{}\biggr(\frac{I_l^{\textrm{sim}}
+I_l^{\textrm{ori}}}{\sum\limits_{l\in \mathcal{L}_1}(I_l^{\textrm{sim}}
+I_l^{\textrm{ori}})}\frac{I_l^{\textrm{sim}}}{I_l^{\textrm{ori}}}\biggr)
\end{align}

$S_1$ is the ratio between the summation of the link weights of the simulated cascades
and that of the original cascades. If $S_1$ is close to 1.0 the links of the original and simulated cascades have almost the same spreading capacity.

$S_2$ and $S_3$ are used to indicate if the shared links play the major role among all links for the original and simulated cascades. If they are near 1.0 it means that
the shared links dominate and thus the simulated cascades are similar to the original
cascades.

$S_4$ indicate the similarity between the overall spreading capacity of the shared 
links of the simulated cascades and that of the original cascades.
$S_4\simeq 1$ will suggest that the overall spreading capacity of the shared links
for the simulated cascades are close to that of the original cascades.

But even when $S_4\simeq 1$ it is still possible that the weight of 
the same link for the original and simulated cascades can be quite different.
Thus $S_5$ is defined to show if the same link is close to each other.
When $S_5$ is near 1.0 it indicates that at least the most important links of the simulated cascades have spreading capacity close to their counterparts for the original cascades.

Note that the similarity indices defined here can not only be used to compare the similarity of the links obtained from original and simulated cascades but can also be used to compare any two sets of links.
For example, we can use them to compare the links from different number of original cascades, which can be denoted by $I_l(M_u^1)$ and $I_l(M_u^2)$ respectively for $M_u^1$ and $M_u^2$ original cascades. 
Since $I_l$ depends on the number of cascades that are used to quantify the interactions, 
$I_l(M_u^1)$ and $I_l(M_u^2)$ should first be normalized before they are compared.
One simple way to normalize them is to divide $I_l(M_u^1)$ by $M_u^1/M_u^2$ or to divide $I_l(M_u^2)$ by $M_u^2/M_u^1$.

\section{Number of Cascades Needed} \label{num needed}

In the above sections we assume there are a total of $M$ cascades and in section \ref{model}
we use $M_u$ of them to generate the tripping probability 
of each component in generation 0 and the interaction matrix. But two questions remain unanswered, which are
how many cascades we need to obtain almost all the interactions between cascading outages and 
how many cascades we need to obtain the dominant interactions that can be used to generate cascades
matching well enough with the original cascades.
In this section we discuss these two questions and determine the lower bounds $M^{\textrm{min}}$ and $M_u^{\textrm{min}}$ respectively for $M$ and $M_u$.

\subsection{Determining Lower Bound for $M$}

More original cascades tend to contain more information about the property of cascading failures of a system, or more specifically the interactions between cascading outages of the components. The added information brought from the added cascades will make the number of identified links increase. However, the number of links will not always grow with the increase of the number of cascades but will saturate after the number of cascades 
is greater than some number $M^{\textrm{min}}$, which can be determined by gradually increasing the number of cascades, recording the number of identified links, and finding the smallest number of cascades that can lead to the saturated number of links.

Assume there are a total of $N_M$ different $M$ ranging from very small number to very large number, 
which are denoted by $M_i,\,i=1,2,\cdots,N_M$.
The number of links for $M_i$ cascades is denoted by $\textrm{card}(\mathcal{L}(M_i))$ 
where $\textrm{card}(\cdot)$ denotes the cardinality of a set, which is a measure of the number of elements of the set.

For $i=1,2,\cdots,N_M-2$ we define 
\begin{equation}
\sigma_i=\sigma(\textrm{card}(\mathcal{L})_i)
\end{equation}
where 
\begin{equation}
\textrm{card}(\mathcal{L})_i=\mathlarger{[}\textrm{card}(\mathcal{L}(M_i)) \; \cdots \; \textrm{card}(\mathcal{L}(M_{N_M}))\mathlarger{]}
\end{equation}
and $\sigma(\cdot)$ is the standard deviation of a vector.

The $\sigma_i$ for $i=N_M-1$ and $i=N_M$ are not calculated since we would like to calculate the standard deviation for at least 3 data points.
Very small and slightly fluctuating $\sigma_i$ can be used to indicate that the number of links begins to saturate after $M_i$ and thus this $M_i$ is determined as $M^{\textrm{min}}$.

The $M^{\textrm{min}}$ original cascades can guarantee that the accuracy on statistical values of interest is good and thus can provide a reference solution.

\subsection{Determining Lower Bound for $M_u$} \label{deter Mu}

When we only want to obtain the dominant interactions that can be used to generate cascades
matching well enough with the original cascades, we do not need as many as $M^{\textrm{min}}$ cascades
but only $M_u^{\textrm{min}}$ cascades to make sure that the propagation capacity of the obtained interaction network $\mathcal{G}(\mathcal{C},\mathcal{L})$ (denoted by $PC^{\mathcal{G}}$) is consistent with that of the original cascades (denoted by $PC^{\textrm{ori}}$). 
Here the physical meaning of the propagation capacity is the average value of the number of caused failures in one cascade. In this section we propose a method to determine $M_u^{\textrm{min}}$.

Since both $PC^{\textrm{ori}}$ and $PC^{\mathcal{G}}$ vary with $M_u$ we denote them by $PC^{\textrm{ori}}(M_u)$ and $PC^{\mathcal{G}}(M_u)$.
$PC^{\textrm{ori}}(M_u)$ can be directly obtained from the original cascades by calculating the average value of the number of failures in generation 1 and the following generations as
\begin{equation}
PC^{\textrm{ori}}(M_u)=\frac{\sum\limits_{m=1}^{M_u}\sum\limits_{g=1}^{\infty}\textrm{card}(F_g^{(m)})}{M_u}.
\end{equation}

In section \ref{keyLink} we define an index $I_l$ for each link $l:\,i\rightarrow j$ which is 
the expected value of the number of failures that are caused through link $l$.
In order to get all the failures that are caused through the link $l:\,i\rightarrow j$ we 
set the number of failures of its source vertex $i$ as $N_i$, which is the total number of failures 
among all the original cascades. Note that $N_i$ contains not only the failures in generation 0, which serve as trigger of cascading failures, but also the failures caused by other component failures.
In this section, however, we only would like to calculate the expected value of the number of failures
caused through link $l:\,i\rightarrow j$ by its source vertex $i$ as generation 0 failures. 
In this case we need to set $N_i^s$ to be $N_{i,0}$, which is the number of failures of component $i$ in generation 0 among all $M_u$ original cascades.
Correspondingly the calculated link index is denoted by $I'_l$ to distinguish with $I_l$ in section \ref{keyLink} for which $N_i^s=N_i$. 
Then the propagation capacity of the interaction network can be written as
\begin{equation}
PC^{\mathcal{G}}(M_u)=\frac{\sum\limits_{l\in \mathcal{L}}I'_l(M_u)}{M_u}.
\end{equation}

When $M_u$ is not large enough it is expected that there will be a big mismatch between
$PC^{\textrm{ori}}(M_u)$ and $PC^{\mathcal{G}}(M_u)$, indicating that the quantified interactions between cascading outages cannot well capture the property of the cascading failure propagation. 
But with the increase of $M_u$ more information will be obtained and thus the mismatch will gradually decrease.
Based on this we increase $M_u$ gradually and identify $M_u^{\textrm{min}}$ as the smallest value that satisfies the following condition
\begin{equation} \label{condition}
|\Delta_{PC}(M_u)| \le \epsilon_{PC} PC^{\textrm{ori}}(M_u)
\end{equation}
where $\Delta_{PC}(M_u)=PC^{\mathcal{G}}(M_u)-PC^{\textrm{ori}}(M_u)$ and $\epsilon_{PC}$ is used to determine the acceptable mismatch.

In order to get $M_u^{\textrm{min}}$ we start from very small $M_u$, such as $M_u^0=100$, and 
calculate the mismatch $\Delta_{PC}(M_u)$. If the condition in (\ref{condition}) is not satisfied $M_u$ is 
increased by a big step $\Delta M_1$ and recalculate $\Delta_{PC}(M_u)$ with the new $M_u$; otherwise $M_u$ is decreased by a small step $\Delta M_2$ until the last $M_u$ for which the condition in (\ref{condition}) is still satisfied.

Note that the number of unnecessary original cascade simulation runs $M_{un}$ is always less than $\Delta M_1$ and actually can be determined by $M_{un}=N_{\Delta M_2} \Delta M_2$ where $N_{\Delta M_2}$ is the number of times $\Delta M_2$ is used for iterations. Also the obtained $M_u^{\textrm{min}}$ is not greater than the smallest possible $M_u$ by $\Delta M_2$.
By decreasing $\Delta M_1$ we can decrease the upper bound of $M_{un}$ but cannot necessarily decrease $M_{un}$.
Differently, by decreasing $\Delta M_1$ we can surely increase the accuracy of the obtained $M_u^{\textrm{min}}$. 
But smaller $\Delta M_1$ or $\Delta M_2$ will increase the time for getting the interaction network and quantifying $I'_l$ for the links.
The selection of $\Delta M_1$ and $\Delta M_2$ can be guided by $\epsilon_{PC}$. Smaller mismatch $\epsilon_{PC}$ will need greater $M_u^{\textrm{min}}$ and thus larger $\Delta M_1$ and $\Delta M_2$ can be chosen to avoid too many times of calculating the interaction network and quantifying $I'_l$.

\section{Cascading Failure Mitigation Measures} \label{miti}

Since the system-level failures of a complex system are actually caused by the interaction of components, 
one possible mitigation measure can be preformed by weakening some key links between component failures, 
which will possibly stop the propagation of cascading failures at an initial stage.
Here weakening key links means reducing the corresponding elements in the interaction matrix $\boldsymbol{B}$.

After validating the proposed interaction model in section \ref{model}, 
this model can be applied to study how the interactions between component failures 
influence the cascading failure risk and to efficiently validate the effectiveness of 
the mitigation measures based on the weakening of key links.

In real systems the weakening of key links can be implemented by blocking some specific protective relays. 
The zone 3 relay blocking method called adaptive distance relay scheme has been discussed in [\ref{CCLiu}].
In this paper relays are blocked under the condition of the tripping of the lines corresponding to the source vertices of the key links. 
Since the key links can cause tremendous expected number of failures and thus play crucial roles in the propagation of cascading failures it should be beneficial to the overall security of the system to stop the propagation from the source vertices of key links to the destination vertices by blocking the operation of the relays of the destination vertices, thus securing time for the operators to take remedial actions, such as re-dispatching the generation or even shedding some loads, and finally helping mitigate catastrophic failures.

This relay blocking strategy under the condition of some specific line tripping can be 
considered as a wide area protection scheme, which can be simulated in AC OPA model by adding a relay blocking module. Note that overloaded lines can be tripped not only by zone 3 relay but can also by other causes, such as tree flashover. The tripping of some lines in two significant outages in the western US in 1996 [\ref{nerc}], the US-Canada Blackout on August 14, 2003 [\ref{us blackout}], and the outage in Italy on September 28, 2003 [\ref{03 blackout}] can all be attributed to tree flashover to some extent, which has been discussed in [\ref{slow}] and [\ref{pra}].

In order to simulate the implementation of the mitigation strategy by weakening some links in real systems and also to compare with the results from the interaction model we add a relay-blocking module in AC OPA model which decreases the tripping probability of some overloaded lines and thus simulates the the weakening of key links. When the line corresponding to the source vertex of a key link is tripped and further causes the overloading of the line corresponding to its destination vertex, the destination vertex line will be tripped with a reduced probability to simulate the part of the role played by blocking of its relay in preventing the line tripping. Due to the reduced tripping probability AC OPA will probably go to its next inner iteration without tripping this destination vertex line and thus AC OPF will be calculated and generation re-dispatching and load shedding will be performed to eliminate the overloading of the destination vertex line.

\section{Results} \label{simulation}

This section presents results for interaction matrix, interaction network, and interaction model.
The cascading outage data is produced by open-loop AC OPA simulation [\ref{AC OPA}], [\ref{AC OPA1}] on IEEE 118-bus system,
which is standard except that the line flow limits are determined with the same method in [\ref{bp13}].
The probability for initial line outage is $p_{0}=0.0001$ and the load variability $\gamma=1.67$, which are the same as [\ref{bp10}], [\ref{bp13}].

AC OPA is a variant of the basic OPA [\ref{opa1}]--[\ref{opa4}].
Basic OPA use DC power flow, for which only active power is considered and the bus voltages are assumed constant. 
In contrast, AC OPA uses AC power flow and thus can consider reactive power and voltage. 
The operation mode of the system is first determined by AC OPF and load shedding and will be readjusted by AC OPF 
until there is no further outage or failure once outages happen. 
Both reactive power and voltage constraints are taken into account in AC OPF.

For testing the interaction quantifying method and the proposed interaction model, 
AC OPA simulation at base case load level is run so as to produce $M$ cascading outages with a nonzero number of line outages.

Note that we only take generating original cascades by AC OPA on IEEE 118-bus system as an example.
AC OPA is only one of the many models that can be used to generate the original cascades needed by the proposed interaction model. 
We can also generate original cascades with basic OPA, improved OPA [\ref{improved}], or OPA with slow process [\ref{slow}].
The test systems can also be real system models used in previous literature, such as the 568-bus Northeast Power Grid of China [\ref{bp13}], [\ref{improved}], [\ref{slow}] or the 1553-bus WECC system [\ref{opa4}]. 

Moreover, the original cascades can also come from utility data, such as the Transmission Availability Data System (TADS), which has 10512 outages recorded by North America utility and can be grouped into 6316 cascades based on outages' timing [\ref{bp12}]. Although the TADS data have been successfully used to study the propagation of cascading failures by using the branching process model in [\ref{bp12}] it is still an open question whether or not the statistical data from utilities are sufficient for obtaining the interaction matrix and interaction network discussed in this paper. One way to check this is to apply the method proposed in section \ref{deter Mu} to determine the minimum number of cascades that are needed to obtain the dominant interactions, which can further be used to generate cascades matching well enough with the original cascades. This definitely deserves more careful discussion in our future work.

Both AC OPA and the interaction model are implemented with Matlab and all tests are carried out on a 3.4 GHz Intel(R) Core(TM) i7-3770 based desktop.

\subsection{Number of Cascades Needed} \label{cascade needed results}

\begin{figure}[!t]
\captionsetup{justification=raggedright,singlelinecheck=false}
\centering
\includegraphics[width=2.8in]{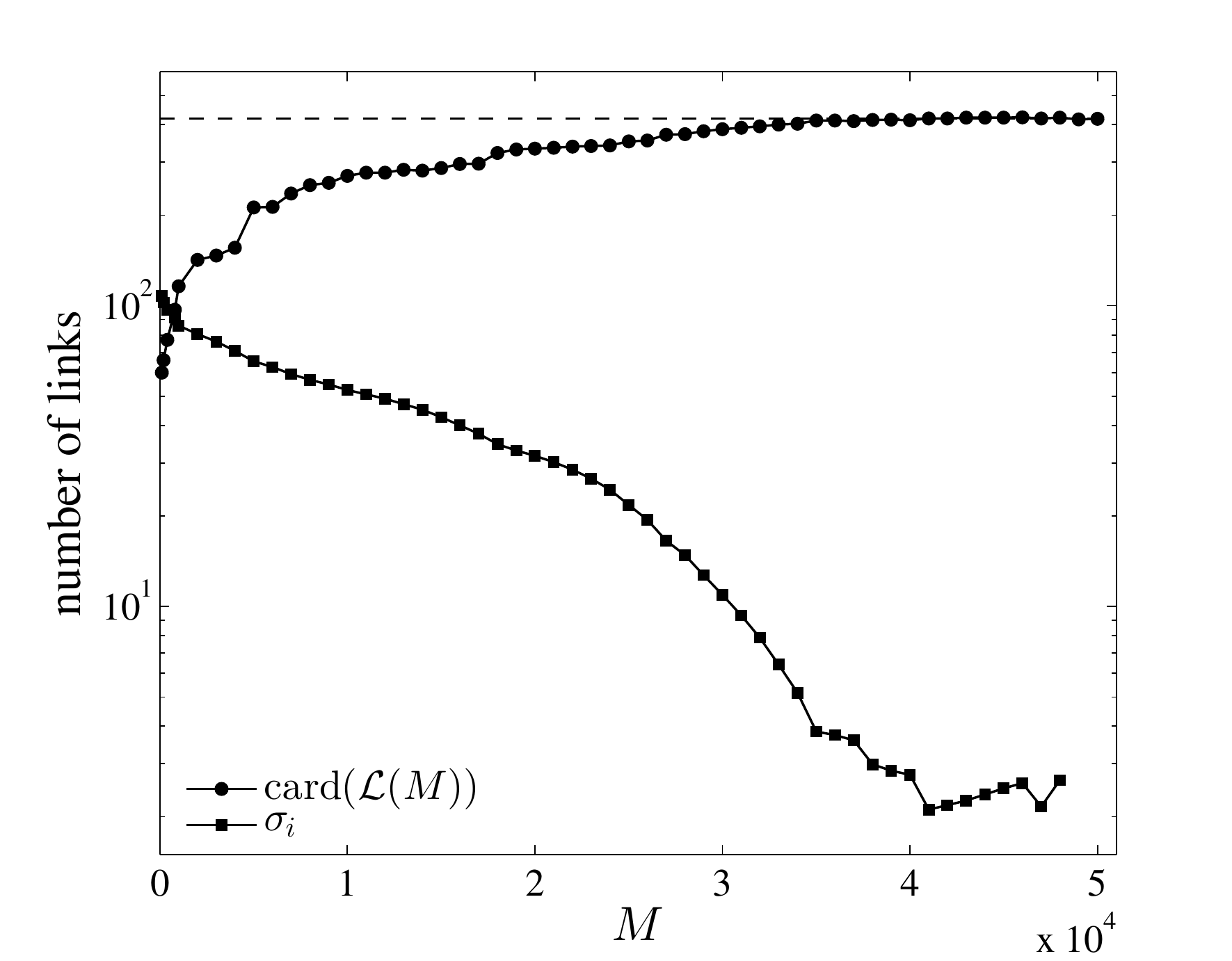}
\caption{Number of links for different $M$. The dash horizontal line indicates $\textrm{card}(\mathcal{L}(M))=419$. }
\label{M}
\end{figure}

In this section we determine the lower bounds for $M$ and $M_u$ by using the method in section \ref{num needed}.
The number of identified links $\textrm{card}(\mathcal{L}(M))$ for different $M$ ranging from 100 to 50000 is shown in Fig. \ref{M}.
We can see that the number of links first grows with the increase of $M$ and finally saturate and only fluctuate slightly when $M$ is large enough. 
There are a total of $N_M=54$ different $M$.
In Fig. \ref{M} we also show $\sigma_i$ for $1 \le i \le N_M-2$.
It is very clear that $\sigma_i$ decreases with the increase of $M$ and finally stabilizes at $M=41000$. Therefore, we choose $M^{\textrm{min}}=41000$ for which the number of identified links is 419. 
The largest number of links for $\textrm{card}(\mathcal{L}(M_i)), i=1,2,\cdots,N_M$ is 423. 
The number of identified links for $M^{\textrm{min}}=41000$ is greater than 99\% of the largest number of links. 
In the rest of this paper we will simulate 41000 cascades for both the AC OPA and the interaction simulation.

In order to determine $M_u^{\textrm{min}}$ we choose $M_u^0$, $\epsilon_{PC}$, $\Delta M_1$, and $\Delta M_2$ in section \ref{deter Mu} as 100, 0.01, 1000, and 100. By setting $\Delta M_2$ to be 100 we can guarantee that the obtained $M_u^{\textrm{min}}$ is not greater than the smallest possible $M_u$ by 100.
After nine $\Delta M_1$ iterations and one $\Delta M_2$ iteration $M_u^{\textrm{min}}$ is determined as 8000, which accounts for 19.51\% of $M^{\textrm{min}}$. The number of unnecessary original cascade simulation runs $M_{un}$ in this case is only 100.
If we set $\epsilon_{PC}$ as 0.05 and correspondingly decrease $\Delta M_1$ and $\Delta M_2$ to 200 and 10, $M_u^{\textrm{min}}$ will be determined as 3680, which only accounts for 8.98\% of $M^{\textrm{min}}$, after nineteen $\Delta M_1$ iterations and two $\Delta M_2$ iterations. 
The number of unnecessary original cascade simulation runs $M_{un}$ in this case is only 20.

\begin{figure}[!t]
\captionsetup{justification=raggedright,singlelinecheck=false}
\centering
\includegraphics[width=2.8in]{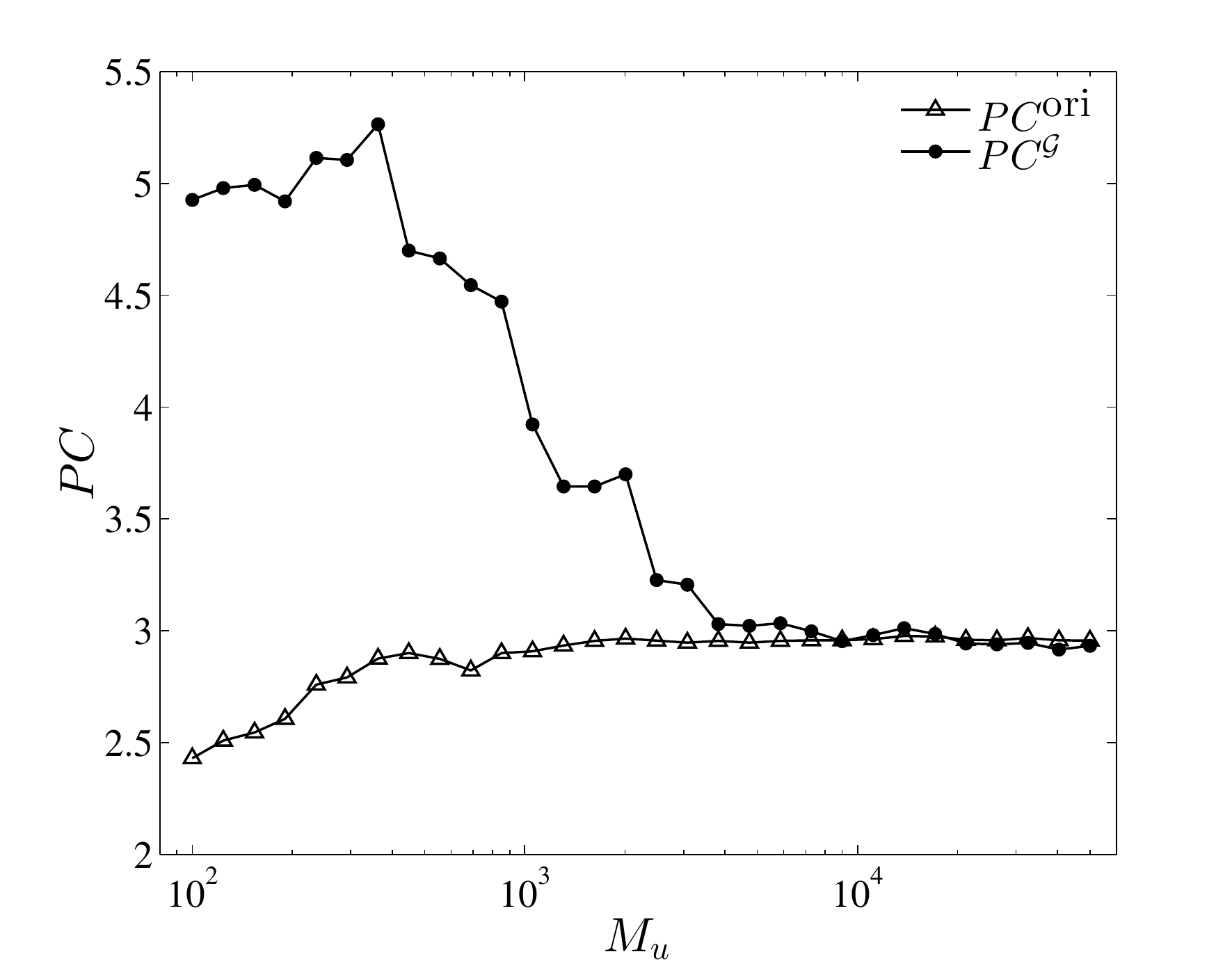}
\caption{Propagation capacity for different $M_u$.}
\label{Mu}
\end{figure}

When we have already generated a large number of cascades we can show how the propagation capacity obtained from original cascades and the interaction network changes with the increase of $M_u$.
This is shown in Fig. \ref{Mu}, in which we can clearly see the trend of the decreasing mismatch between $PC^{\textrm{ori}}(M_u)$ and $PC^{\mathcal{G}}(M_u)$.

In Fig. \ref{Mu} we can also see that the quantified interaction network tends to obtain an overestimated propagation capacity for a small number of $M_u$. As has been tentatively discussed in section \ref{interaction}, this is because for any component $j$ no component has much greater $a_{ij}$ than the others when only using a small number of $M_u$ and thus several components have to be determined as the cause. This further leads to the overestimation of the propagation of cascading failures.

In order to show this we analyze how $\boldsymbol{A}$ changes with the increase of $M_u$ in more detail.
Let $\mathcal{C}^{\textrm{caused}}$ denote the set of components that are caused by other components in $\boldsymbol{A}$, which actually corresponds to the columns of $\boldsymbol{A}$ with nonzero elements. For any $j\in \mathcal{C}^{\textrm{caused}}$, if $\textrm{card}(\{i|a_{ij}>0\})>1$ and $\max\limits_{i\in \mathcal{C}} a_{ij}=\bar{a}_j$ where $\mathcal{C}$ is all the components and $\bar{a}_j$ is the average value for all $a_{ij}>0$ then component $j$ is called completely cause-indistinguishable. 
The set of completely cause-indistinguishable components is denoted by $\mathcal{C}^{id}$.
Actually when the given condition holds it is easy to prove by contradiction that all nonzero $a_{ij}$ will be equal. Thus in this case it is completely impossible to distinguish which cause of component $j$ is more possible than the others. 
The ratio of completely cause-indistinguishable components among all caused components can be calculated as
\begin{equation}
r_{id}=\frac{\textrm{card}(\mathcal{C}^{id})}{\textrm{card}(\mathcal{C}^{\textrm{caused}})}.
\end{equation}

\begin{figure}[!t]
\captionsetup{justification=raggedright,singlelinecheck=false}
\centering
\includegraphics[width=2.8in]{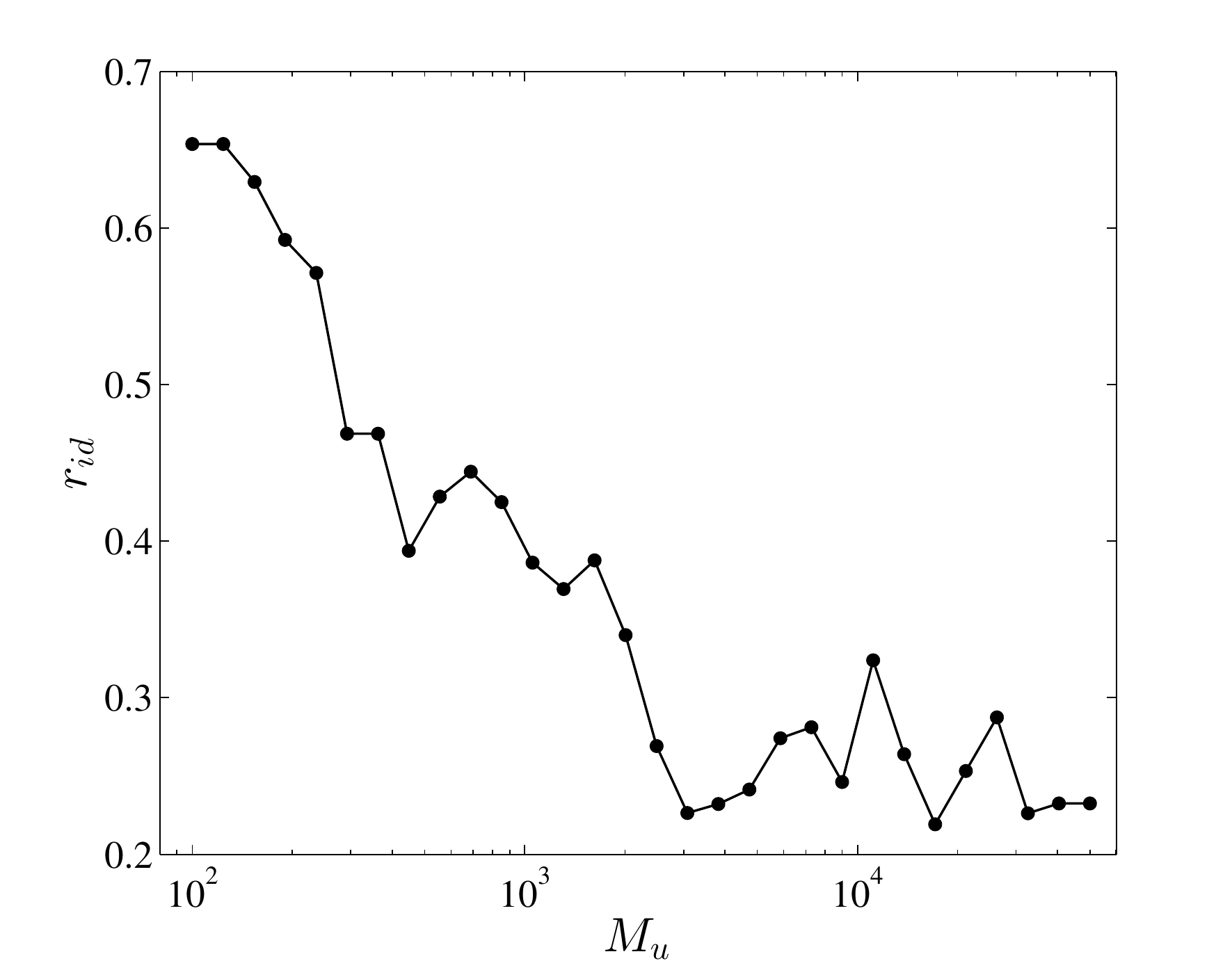}
\caption{Ratio of completely cause-indistinguishable components.}
\label{rid}
\end{figure}

We show $r_{id}$ in Fig. \ref{rid} and it is clearly seen that the completely cause-indistinguishable components account for a large proportion when $M_u$ is small and will gradually decrease to a relatively low level with the increase of $M_u$. The high ratio of the completely cause-indistinguishable components for a small number of $M_u$ leads to the overestimated interactions between component failures and further explains the too great propagation capacity of the interaction network.

\subsection{Interaction Matrix and Interaction Network} \label{interactionMatrix}

There are 186 lines in IEEE 118-bus system and thus $\boldsymbol{B}$ is a 186 $\times$ 186 square matrix.
Table \ref{nonzero} shows the number of components $n$, 
the number of cascades used to quantify the interactions $M_u$, 
the number of links $\textrm{card}(\mathcal{L})$, which is also the number of $\boldsymbol{B}$'s nonzero elements,
and the ratio of nonzero elements $r=\textrm{card}(\mathcal{L})/{n^2}$.
It is seen that $r$ is very small, indicating that the interaction matrix is very sparse and 
that only a small fraction of lines interact with each other.

\begin{table}[!t]
\renewcommand{\arraystretch}{1.3}
\captionsetup{labelsep=space,font={footnotesize,sc}}
\caption{\\Nonzero Elements in $\boldsymbol{B}$ for IEEE 118-Bus System}
\label{nonzero}
\centering
\begin{tabular}{ccccc}
\hline
 model & $n$ & $M_u$ & $\textrm{card}(\mathcal{L})$ & $r$ \\
\hline
\scriptsize AC OPA & 186 & 41000 & 419 & 0.0121  \\
\scriptsize AC OPA & 186 & 8000 & 252 & 0.00730  \\
\scriptsize AC OPA & 186 & 3680 & 156 & 0.00450  \\
\hline
\end{tabular}
\end{table}

\begin{figure}[!t]
\captionsetup{justification=raggedright,singlelinecheck=false}
\centering
\includegraphics[width=2.7in]{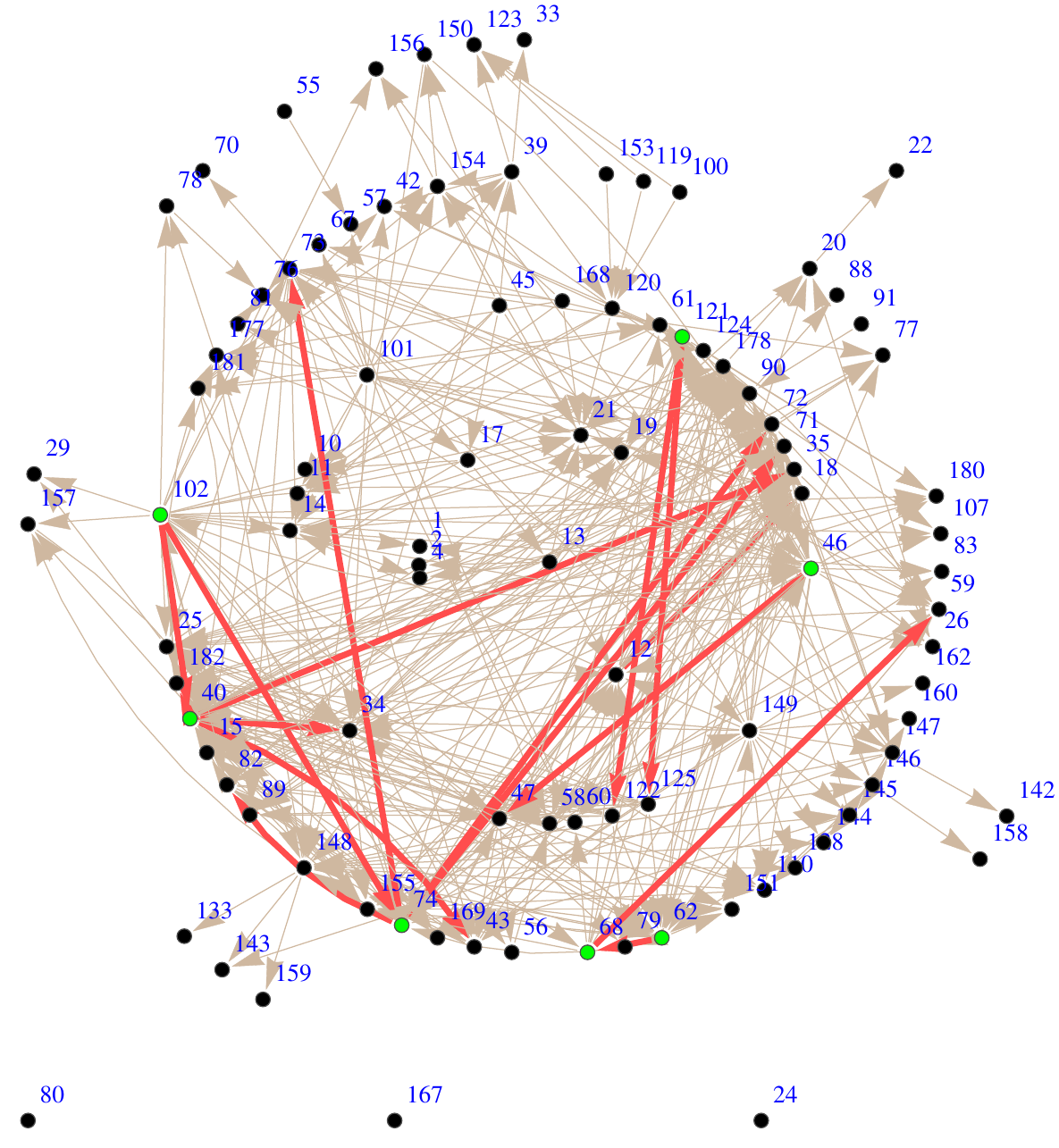}
\caption{Interaction network for using 41000 original cascades.}
\label{link118_41000}
\end{figure}

\begin{figure}[!t]
\captionsetup{justification=raggedright,singlelinecheck=false}
\centering
\includegraphics[width=2.7in]{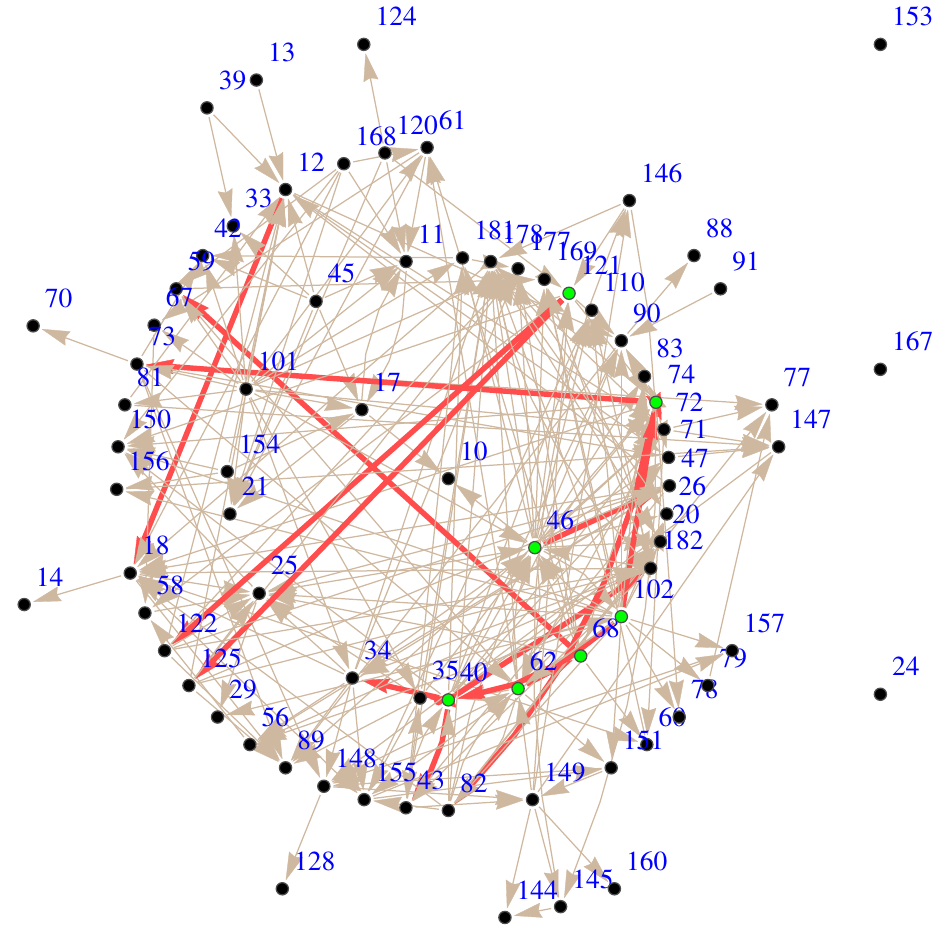}
\caption{Interaction network for using 8000 original cascades.}
\label{link118_8000}
\end{figure}

\begin{figure}[!t]
\captionsetup{justification=raggedright,singlelinecheck=false}
\centering
\includegraphics[width=2.7in]{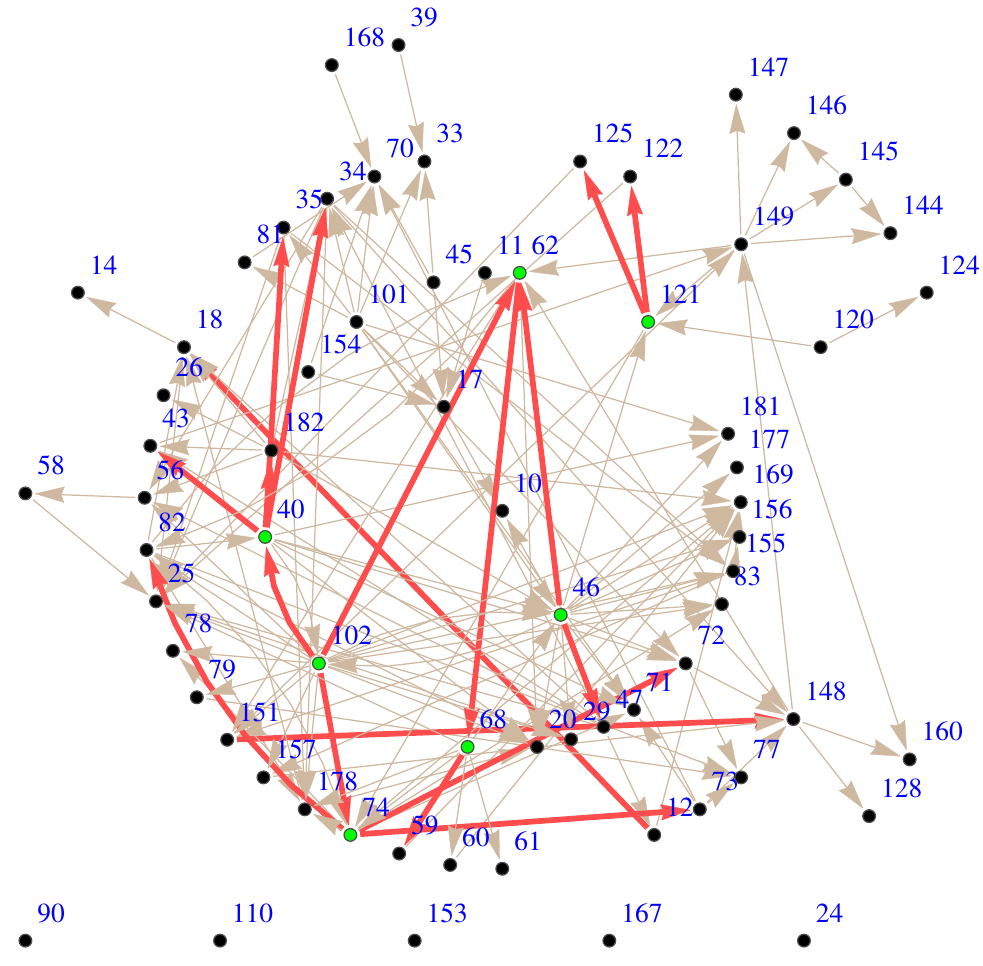}
\caption{Interaction network for using 3680 original cascades.}
\label{link118_3680}
\end{figure}

The corresponding directed weighted interaction network is shown in Figs. \ref{link118_41000}--\ref{link118_3680}, 
in which the dots denote lines in IEEE 118-bus system and the arrows denote the links between lines.
Here we do not show the weights of the links but only the topology of the interaction network.
This network is different from the one-line diagram of IEEE 118-bus system, for which the vertices are buses and the undirected links between vertices are lines.

We compare the links obtained from 8000 and 3680 original cascades 
with the reference 41000 original cascades by calculating the similarity indices 
defined in section \ref{validate}, which are shown in Table \ref{index}.
As is discussed in section \ref{validate}, since $I_l$ depends on the number of cascades that are used
the index $I_l$ of the links should first be normalized before calculating the similarity indices.
Specifically, we divide the $I_l$ for 41000 cascades by $41000/8000$ and $41000/3680$.
Results in Table \ref{index} shows that the first four indices are all close to 1 and thus indicate the links from smaller number of 
cascades are similar to those for more cascades. But the fifth index $S_5$, which requires more strict similarity, 
is not so close to 1, indicating that some information is missing for fewer original cascades.
However, we will show in section \ref{validation} that the links obtained from simulated more cascades with the proposed interaction model can be very similar to those from 41000 original cascades.

\begin{table}[H]
\renewcommand{\arraystretch}{1.3}
\captionsetup{labelsep=space,font={footnotesize,sc}}
\caption{\\Similarity Indices for 8000 and 3680 Original Cascades Compared with 41000 Original Cascades}
\label{index}
\centering
\begin{tabular}{ccccccc}
\hline
model & $M_u$ & $S_1$ & $S_2$ & $S_3$ & $S_4$ & $S_5$ \\
\hline
\scriptsize AC OPA & 8000 & 1.02 & 0.996 & 0.991 & 1.02 & 2.79 \\
\scriptsize AC OPA & 3680 & 1.04 & 0.987 & 0.985 & 1.04 & 6.41 \\
\hline
\end{tabular}
\end{table}

\subsection{Key Link and Key Component Identification} \label{identification}

In this section key links and key components that play important roles in the propagation of cascading failures 
are identified by using the method in section \ref{keyLink}. 
Both $\epsilon_l$ and $\epsilon_s$ are chosen as 0.15 (slightly greater than 1/10) to make the weights of all key links and all key components are in the same order.

The identified key links, which are actually line pairs in IEEE 118-bus system, 
and their weights $I_l$ for the three cases separately using 41000, 8000, and 3680 original cascades are listed in Table \ref{linePairs}.
They are also shown in Figs. \ref{link118_41000}--\ref{link118_3680} by red arrows.
The numbers in the parentheses for $I_l(8000)$ and $I_l(3680)$ are the ranking of the key links.
We can see that the identified key links for the three cases are almost the same and the ranking of the links is also quite similar. The link $(53,\,54)\rightarrow (49,\,51)$ is only identified for using 41000 cascades and the links $(65,\,66)\rightarrow (45,\,46)$, $(35,\,36)\rightarrow (45,\,46)$, and $(80,\,97)\rightarrow (80,\,96)$ are only identified for using 3680 cascades.
Although the link $(53,\,54)\rightarrow (49,\,51)$ is not identified for using 8000 and 3680 cascades,
the corresponding $I_l$ for the two cases separately rank the 16th and 19th and are equal to 378 and 157, which are much greater than the average values of all link indices 109 and 82.
For the three cases the number of key links are only 3.82\%, 5.95\%, and 11.54\% of all the links but the summation of their weights accounts for 88.08\%, 85.12\%, and 87.65\% of the total weights of all links.

\begin{table}[!t]
\renewcommand{\arraystretch}{1.3}
\captionsetup{labelsep=space,font={footnotesize,sc}}
\caption{\\Key Links for IEEE 118-Bus System}
\label{linePairs}
\centering
\begin{tabular}{ccccc}
\hline
 $i \rightarrow j$ & line pairs & $I_l(41000)$ & $I_l(8000)$ & $I_l(3680)$ \\
\hline
$74 \rightarrow 72$ & \tabincell{l}{$(53,\,54)\rightarrow$ \\ $(51,\,52)$} & 12582 & 2486 (2) & 1150 (1) \\
$74 \rightarrow 73$ & \tabincell{l}{$(53,\,54)\rightarrow$ \\ $(52,\,53)$} & 12469  & 2535 (1) & 1150 (2) \\
$40 \rightarrow 34$ & \tabincell{l}{$(29,\,31)\rightarrow$ \\ $(27,\,28)$} & 11920 & 2305 (3) & 1031 (3) \\
$40 \rightarrow 35$ & \tabincell{l}{$(29,\,31)\rightarrow$ \\ $(28,\,29)$} & 11421 & 2167 (4) & 974 (5) \\
$74 \rightarrow 82$ & \tabincell{l}{$(53,\,54)\rightarrow$ \\ $(56,\,58)$} & 10802 & 2153 (5) & 1000 (4) \\
$62 \rightarrow 68$ & \tabincell{l}{$(45,\,46)\rightarrow$ \\ $(45,\,49)$} & 9865 & 1915 (6) & 940 (6) \\
$121 \rightarrow 122$ & \tabincell{l}{$(77,\,78)\rightarrow$ \\ $(78,\,79)$} & 9601 & 1912 (7) & 862 (7) \\
$121 \rightarrow 125$ & \tabincell{l}{$(77,\,78)\rightarrow$ \\ $(79,\,80)$} & 9599 & 1912 (8) & 862 (8) \\
$40 \rightarrow 182$ & \tabincell{l}{$(29,\,31)\rightarrow$ \\ $(114,\,115)$} & 6687 & 1261 (9) & 592 (9) \\
$46 \rightarrow 47$ & \tabincell{l}{$(35,\,36)\rightarrow$ \\ $(35,\,37)$} & 5536 & 1092 (10) & 502 (10) \\
$12 \rightarrow 18$ & \tabincell{l}{$(11,\,12)\rightarrow$ \\ $(13,\,15)$} & 5388 & 1001 (11) & 450 (11) \\
$40 \rightarrow 43$ & \tabincell{l}{$(29,\,31)\rightarrow$ \\ $(27,\,32)$} & 4475 & 886 (12) & 404 (12) \\
$68 \rightarrow 59$ & \tabincell{l}{$(45,\,49)\rightarrow$ \\ $(43,\,44)$} & 3690 & 706 (13) & 347 (13) \\
$102 \rightarrow 74$ & \tabincell{l}{$(65,\,66)\rightarrow$ \\ $(53,\,54)$} & 3135 & 635 (14) & 273 (14) \\
$74 \rightarrow 71$ & \tabincell{l}{$(53,\,54)\rightarrow$ \\ $(49,\,51)$} & 1977 & -- & -- \\
$102 \rightarrow 40$ & \tabincell{l}{$(65,\,66)\rightarrow$ \\ $(29,\,31)$} & 1968 & 415 (15) & 178 (15) \\
$102 \rightarrow 62$ & \tabincell{l}{$(65,\,66)\rightarrow$ \\ $(45,\,46)$} & -- & -- & 178 (16) \\
$46 \rightarrow 62$ & \tabincell{l}{$(35,\,36)\rightarrow$ \\ $(45,\,46)$} & -- & -- & 178 (17) \\
$151 \rightarrow 148$ & \tabincell{l}{$(80,\,97)\rightarrow$ \\ $(80,\,96)$} & -- & -- & 175 (18) \\
\hline
\end{tabular}
\end{table}

The identified key components, the corresponding lines, and their out-strengths
for using using 41000, 8000, and 3680 original cascades are listed in Table \ref{keyComponentLine}, in which 
the numbers in the parentheses are the ranking of the key components.
It is seen that the identified key components and their ranking are exactly the same for the three cases.
They are also highlighted in Figs. \ref{link118_41000}--\ref{link118_3680} by green vertices.
The tripping of these lines will cause severe consequences and thus should be prevented to the greatest extent.
For IEEE 118-bus system there are a total of 186 components and among them 97, 76, 68 components are involved in the 41000, 8000, and 3680 original cascades. For the three number of original cascades, the number of key components are 3.76\% of all components and separately 7.22\%, 9.21\%, and 10.29\% of the involved components and the summation of the out-strengths of the key components accounts for 89.12\%, 88.83\%, and 88.65\% of the total out-strengths of the involved components.

\begin{table}[!t]
\renewcommand{\arraystretch}{1.3}
\captionsetup{labelsep=space,font={footnotesize,sc}}
\caption{\\Key Components for IEEE 118-Bus System}
\label{keyComponentLine}
\centering
\begin{tabular}{cccccc}
\hline
key component & line & $s_i^{\textrm{out}}(41000)$ & $s_i^{\textrm{out}}(8000)$ & $s_i^{\textrm{out}}(3680)$ \\
\hline
74 &  $(53,\,54)$ & 37925 & 7577 (1) & 3467 (1) \\
40 &  $(29,\,31)$ & 34613 & 6651 (2) & 3015 (2) \\
121 & $(77,\,78)$ & 19219 & 3833 (3) & 1728 (3) \\
62 &  $(45,\,46)$ & 9892 & 1926 (4) & 944 (4) \\
102 & $(65,\,66)$ & 8039 & 1826 (5) & 892 (5) \\
46 & $(35,\,36)$ & 6645 & 1377 (6) & 734 (6) \\
68 & $(45,\,49)$ & 6210 & 1211 (7) & 594 (7) \\
\hline
\end{tabular}
\end{table}

\begin{figure}[!t]
\captionsetup{justification=raggedright,singlelinecheck=false}
\centering
\includegraphics[width=3.45in]{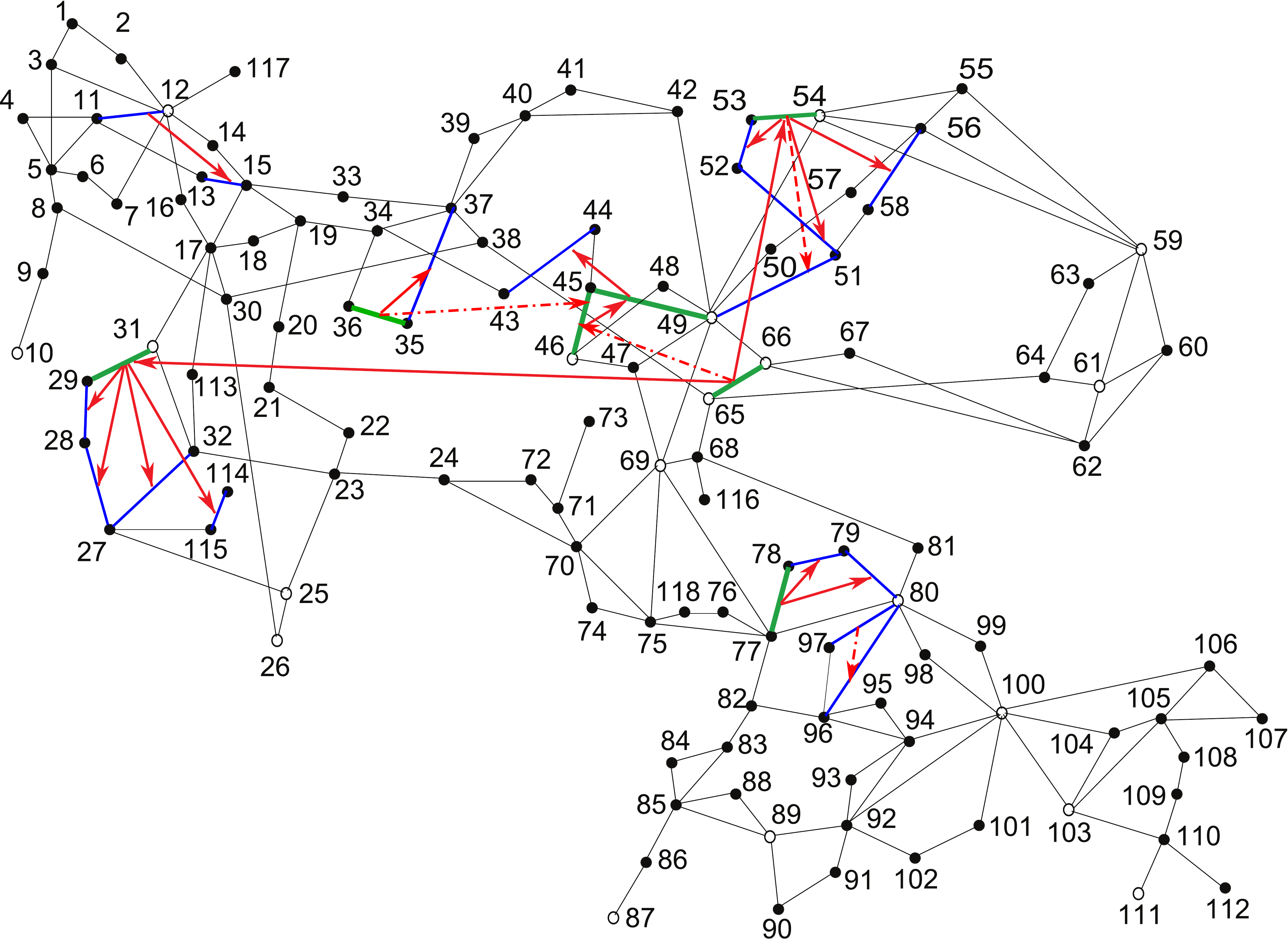}
\caption{Key links and key components for IEEE 118-bus system. Key links shared by the three cases for using 41000, 8000, and 3680 cascades are denoted by red solid arrow; key links only for using 41000 cascades are denoted by red dash arrow; key links only for using 3680 cascades are denoted by red dash-dotted arrow; key components are denoted by green lines and the other lines involved in the key links are denoted by blue lines.}
\label{118bussystem}
\end{figure}

The identified key links and key components are denoted on the one-line diagram of IEEE 118-bus system, 
which is shown in Fig. \ref{118bussystem}. 
It is seen that the lines corresponding to the source and destination vertices of some key links 
can be topologically far away from each other, such as link $(65,\,66)\rightarrow (53,\,54)$ and $(65,\,66)\rightarrow (29,\,31)$,
although for most key links the source and destination vertices are lines that are topologically close to each other, such as link $(53,\,54)\rightarrow (51,\,52)$ and $(53,\,54)\rightarrow (52,\,53)$.
This is because the interactions and links are obtained from simulated cascades 
generated by AC OPA model which not only considers the topology of the power network 
but also other physics of the system, such as power flow and the operator response.
These factors can also make some components tightly coupled.
Therefore, only the topology is not sufficient to fully
reveal the nonlocal property of the propagation of cascading
failures, which has been observed in several blackouts around
the world [\ref{nerc}]--[\ref{03 blackout}], and is thus better to be used for studying cascading failures.

\subsection{Model Validation} \label{validation}

\begin{figure}[!t]
\captionsetup{justification=raggedright,singlelinecheck=false}
\centering
\includegraphics[width=2.8in]{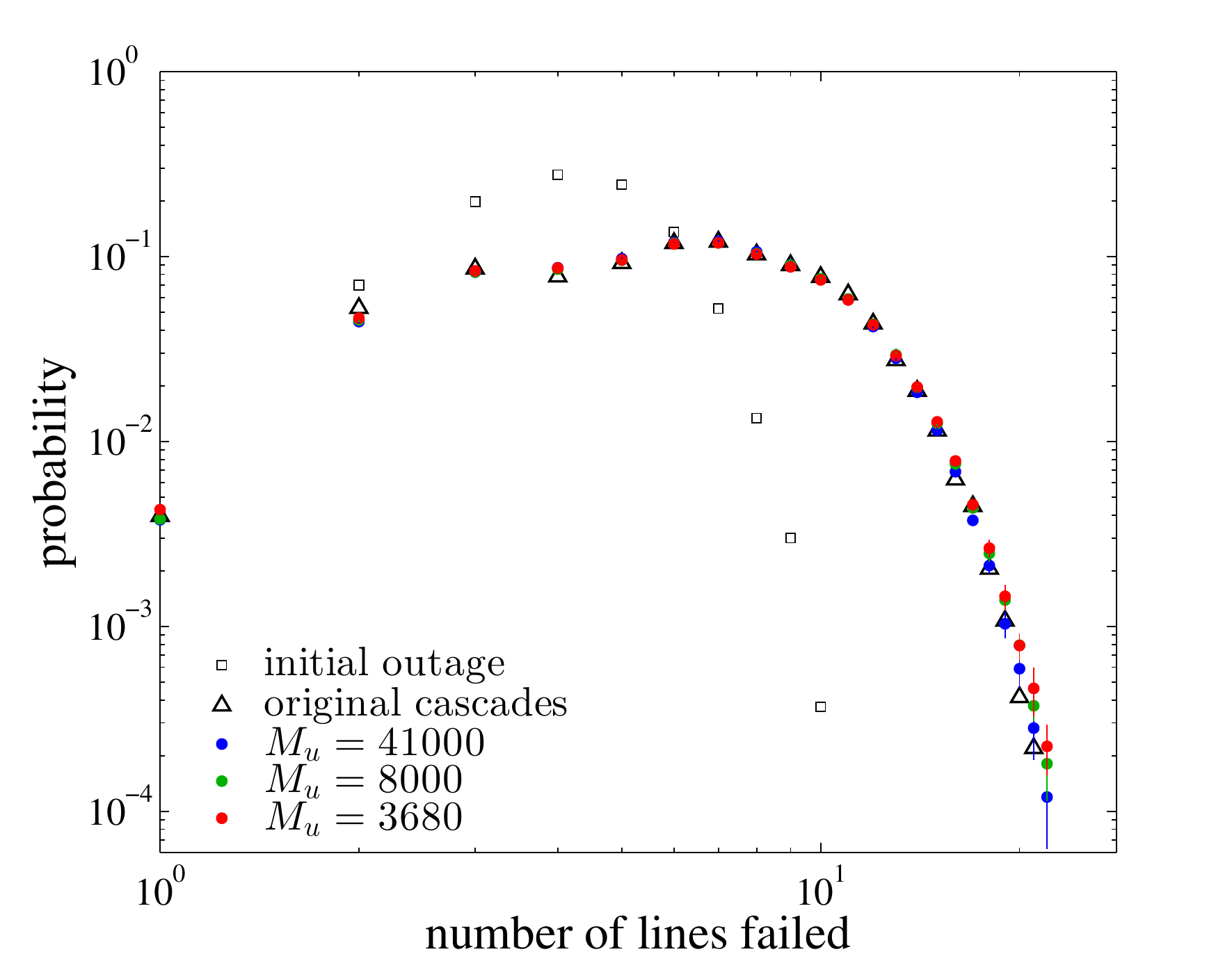}
\caption{Probability distributions of the total number of line outages 
for original and simulated cascades.
Triangles indicate total numbers of line outages of 
the original cascades; blue, green, and red dots separately indicate total numbers of line outages of 
the simulated cascades for using 41000, 8000, and 3680 original cascades to quantify interactions; vertical lines indicate the standard deviations for 20 times of interaction model simulations; squares indicate initial line outages.}
\label{line}
\end{figure}

In this section the proposed interaction model is validated with the four methods
discussed in section \ref{validate}.
The probability distributions of the total number of line outages
for original cascades (triangles) and simulated cascades (dots) by using 41000, 8000, and 3680 original cascades to quantify interactions are shown in Fig.~\ref{line}.
For the simulated cascades we simulate 41000 cascades for 20 times for each case and show their average probability distribution and the standard deviations (vertical lines).

It is seen that the distributions of total line outages of the original and simulated cascades match well and the standard deviations of the probability distributions of the simulated cascades are small. 
This suggests that the proposed interaction model can generate cascades with similar statistical properties to the original cascades.
The dramatic difference between the distributions of the initial and total outages also suggest that 
the cascading failure is able to propagate a lot.
This is because of the interaction between component failures denoted by the sparse interaction matrix. 
If all elements of $\boldsymbol{B}$ are zero and the components do not interact at all,
all cascades will stop immediately after initial line outages 
and the distribution of the total line outages will be the same as initial line outages.
Thus, although being sparse, the interaction matrix does take effect.

To quantitatively compare the original cascades and the simulated cascades,
the branching process is applied to estimate their average propagation. 
The average value $m_{\hat{\lambda}}$ and the standard deviation $\sigma_{\hat{\lambda}}$ of the average propagation for 20 times of simulation are listed in Table \ref{lambda}, in which $M_u$ is the number of original cascades used to quantify the interactions between component failures and $M$ is the number of cascades simulated by the interaction model.
It is seen that the estimated average propagation of the simulated cascades
are very close to that of the original cascades and the standard deviations are very small, 
indicating that the simulated cascades have similar propagation property to the original cascades.

\begin{table}[H]
\renewcommand{\arraystretch}{1.3}
\captionsetup{labelsep=space,font={footnotesize,sc}}
\caption{\\Average Propagation for Original and Simulated Cascades}
\label{lambda}
\centering
\begin{tabular}{ccccc}
\hline
model & $M_u$ & $M$ & $m_{\hat{\lambda}}$ & $\sigma_{\hat{\lambda}}$ \\
\hline
\scriptsize AC OPA & -- & 41000 & 0.402 & -- \\
interaction & 41000 & 41000 & 0.402 & 0.000969 \\
interaction & 8000 & 41000 & 0.410 & 0.000800 \\
interaction & 3680 & 41000 & 0.412 & 0.000899 \\
\hline
\end{tabular}
\end{table}

\begin{figure}
\captionsetup{justification=raggedright,singlelinecheck=false}
\centering
\includegraphics[width=2.8in]{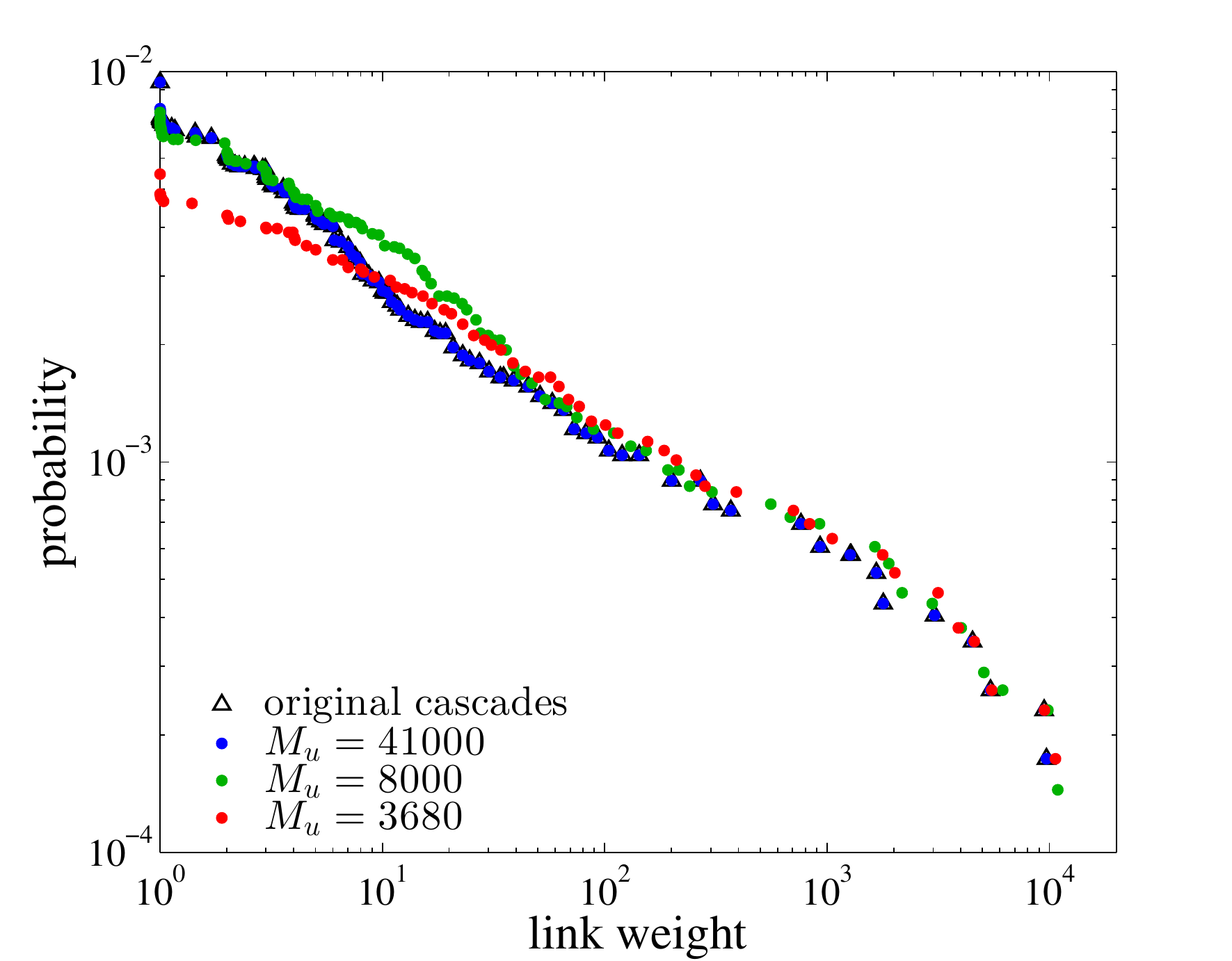}
\caption{CCD of the link weight for original and simulated cascades. Triangles indicate CCD of the link weight of original cascades; blue, green, and red dots separately indicate CCD of the link weight of 
the simulated cascades for using 41000, 8000, and 3680 original cascades to quantify interactions.}
\label{linkWeight}
\end{figure}

The complementary cumulative distributions (CCD) of the link weights 
for the original and simulated cascades for $M_u=41000$, $M_u=8000$, and $M_u=3680$ are shown in Fig. \ref{linkWeight}.
The CCD for simulated cascades are the distribution for all the links obtained from 20 times of simulation.
Note that we group the links with very big weights together and calculate the CCD for their average value 
to avoid the possible unreliable estimation for rarer events 
since the number for each of them can be very small.
Also when we calculate the CCD the zero elements are considered as links with zero weights.
Fig. \ref{linkWeight} shows that the two distributions match very well. 
Both of them follow obvious power law and can range from 1 to more than 10000, 
suggesting that a small number of links can cause much greater consequences than most of the others.

\begin{figure}
\captionsetup{justification=raggedright,singlelinecheck=false}
\centering
\includegraphics[width=2.8in]{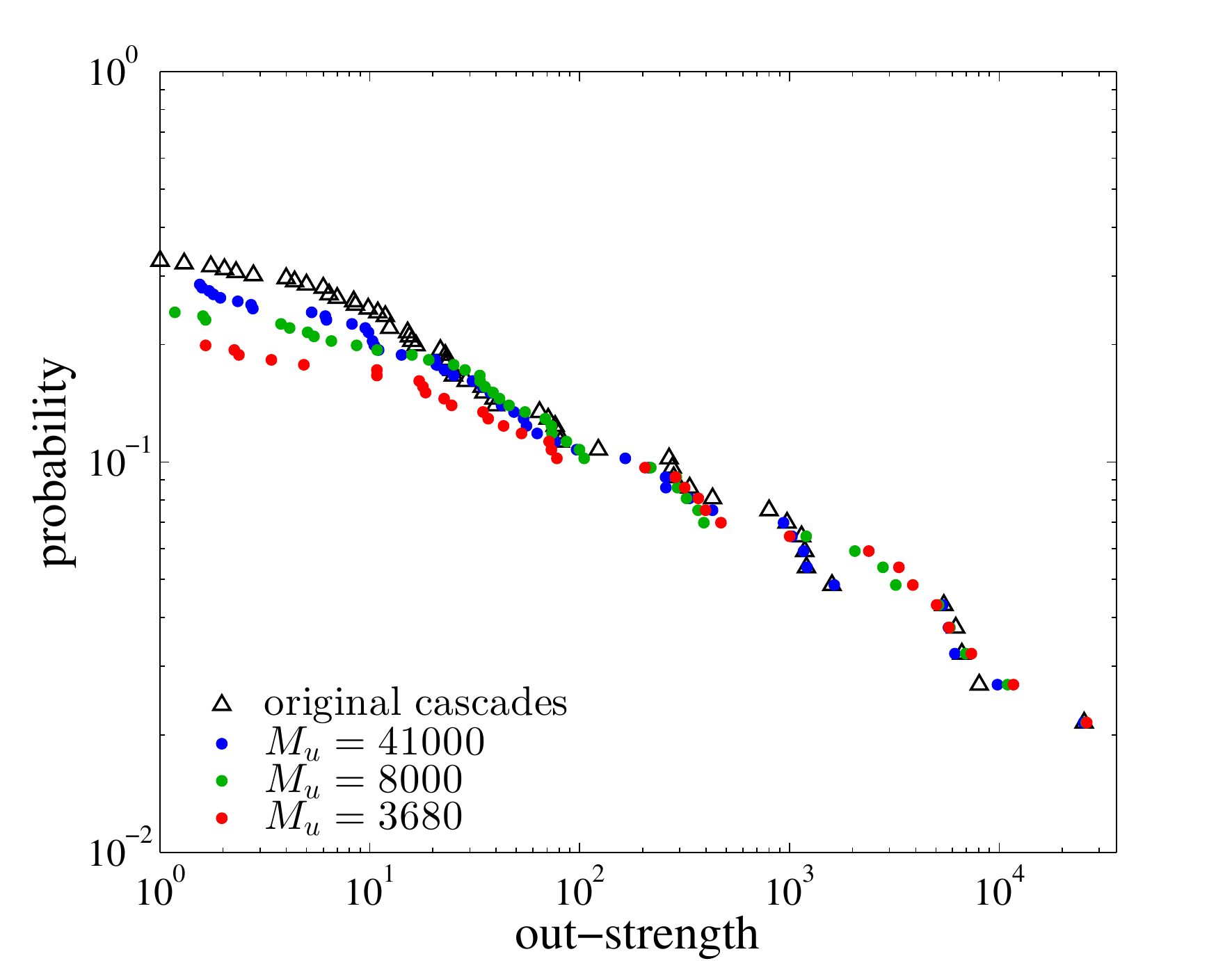}
\caption{CCD of the vertex out-strength for original and simulated cascades.}
\label{outstrength}
\end{figure}

\begin{figure}
\captionsetup{justification=raggedright,singlelinecheck=false}
\centering
\includegraphics[width=2.8in]{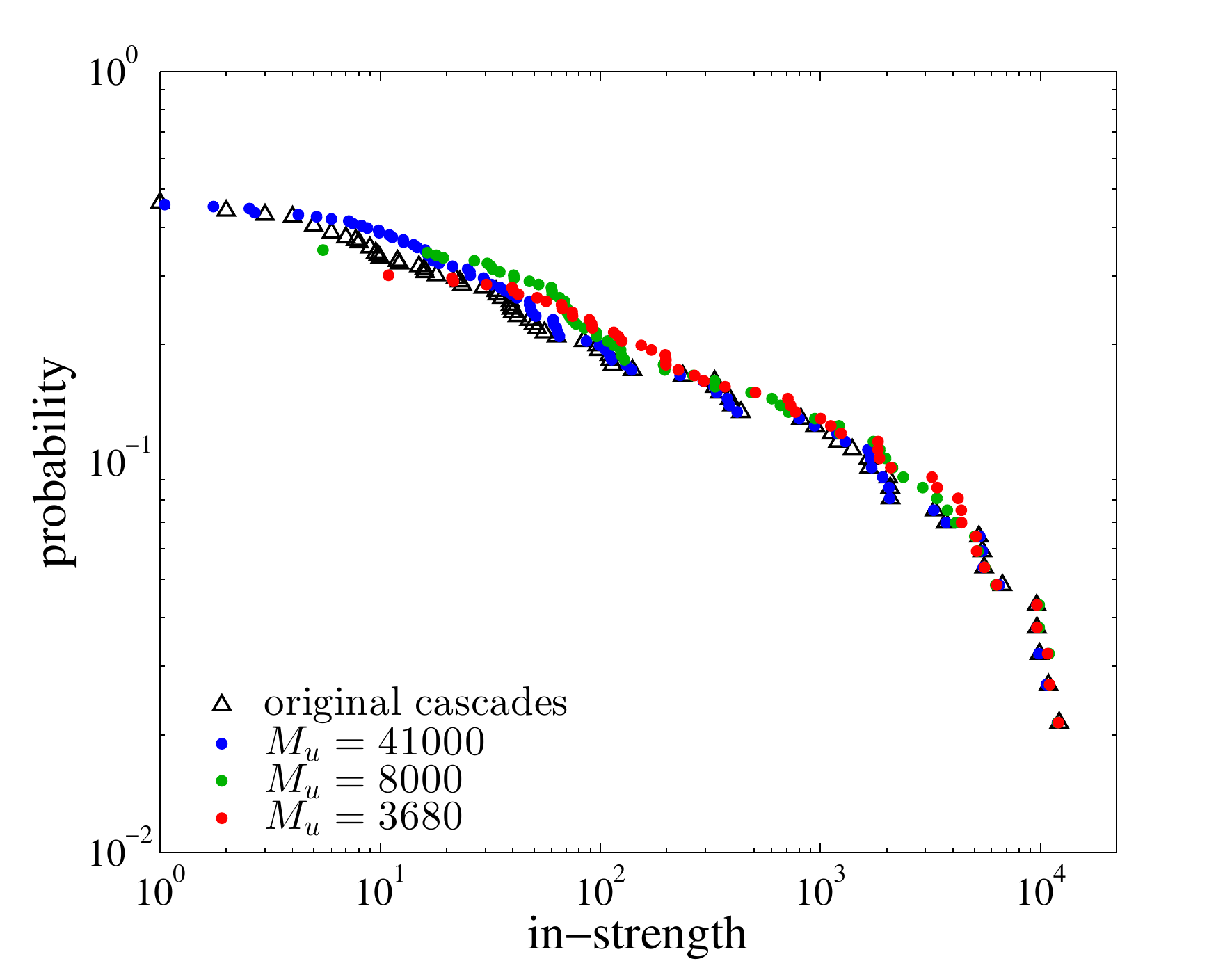}
\caption{CCD of the vertex in-strength for original and simulated cascades.}
\label{instrength}
\end{figure}

The CCD of the vertex out-strength and in-strength for original and simulated cascades
are shown in Figs. \ref{outstrength}--\ref{instrength}.
Similar to link weights, the CCD for simulated cascades are the distribution for all the vertex out-strength and in-strength obtained from 20 times of simulation. 
We also group the vertices with very big out-strength or in-strength together 
and the components that do not appear in the interaction network are considered as 
vertices with zero out-strength and in-strength.
The strength distributions of the original and simulated cascades match very well,
indicating that the simulated cascades share similar features to the original cascades from an overall point of view. An obvious power law behavior can also be seen, which means that the failure of most vertices (components) have small consequences while a small number of them have much greater impact.

The five similarity indices defined in section \ref{validate} 
for the quantified links from the simulated 41000 cascades and the 
41000 original cascades are listed in Table \ref{index1}.
It is seen that all five indices are close to 1.0 and thus the links obtained 
from the original and simulated cascades are actually quite similar.
The standard deviations of the similarity indices for 20 times of simulation
are listed in the parentheses and are all very small.
The similarity indices between the quantified links from 8000 and 3680 original cascades and those 
from 41000 cascades are listed in Table \ref{index}.
By comparing the results in Table \ref{index} and Table \ref{index1} we can see that 
the first four indices are only slightly different but the fifth index significantly improves 
for simulating 41000 cascades with the interaction model by using fewer original cascades to 
quantify interactions, indicating that the missing information for fewer original cascades 
can be recovered to a great extent by simulating more cascades with the interaction model, which has high-level probabilistic property and can reveal more rare events by doing a large number of simulations.
It is also not surprising that using fewer original cascades cannot get as good results as using the whole 41000 original cascades since fewer original cascades will inevitably miss some information to some extent.

\begin{table}[H]
\renewcommand{\arraystretch}{1.3}
\captionsetup{labelsep=space,font={footnotesize,sc}}
\caption{\\Similarity Indices for Original and Simulated Cascades}
\label{index1}
\centering
\begin{tabular}{cccccc}
\hline
$M_u$ & $S_1$ & $S_2$ & $S_3$ & $S_4$ & $S_5$ \\
\hline
41000  & \tabincell{c}{0.994 \\(0.00900)} & \tabincell{c}{0.994 \\ (0.000619)} & \tabincell{c}{0.973 \\ (0.0168)} & \tabincell{c}{0.974 \\ (0.0194)} & \tabincell{c}{1.01 \\ (0.0122)} \\
8000 & \tabincell{c}{1.05 \\(0.00570)} & \tabincell{c}{0.989 \\ (0.00280)} & \tabincell{c}{0.973 \\(0.00930)} & \tabincell{c}{1.04 \\ (0.0146)} & \tabincell{c}{1.14 \\ (0.0379)} \\
3680 & \tabincell{c}{1.07 \\ (0.00540)} & \tabincell{c}{0.987 \\ (0.00240)} & \tabincell{c}{0.944 \\ (0.00180)} & \tabincell{c}{1.03 \\ (0.00690)} & \tabincell{c}{1.17 \\ (0.0207)} \\
\hline
\end{tabular}
\end{table}

\subsection{Cascading Failure Mitigation} \label{mitigationResult}

We assume that 90\% of the tripping of overloaded lines are due to the operation of zone 3 relays. For the interaction model the weakening of the key links is simulated by reducing the
corresponding elements in the interaction matrix by 90\%.
For each of the three cases in which $M_u=41000$, $M_u=8000$, and $M_u=3680$, 
the key links identified in Table \ref{linePairs} of section \ref{identification} are weakened
by reducing the corresponding elements in $\boldsymbol{B}(M_u)$ by 90\%.
By doing this we get $\boldsymbol{B}_{\textrm{int}}(M_u)$.
For comparison the same number of randomly chosen links are also weakened in the same way.
for which $\boldsymbol{B}_{\textrm{rand}}(M_u)$ is obtained.
Cascading failures are separately simulated with the proposed model 
by using $\boldsymbol{B}_{\textrm{int}}(M_u)$ and $\boldsymbol{B}_{\textrm{rand}}(M_u)$.
The two mitigation strategies are respectively called intentional mitigation 
and random mitigation. Each case for each mitigation strategy is simulated for 20 times.

\begin{figure}[!t]
\captionsetup{justification=raggedright,singlelinecheck=false}
\centering
\includegraphics[width=2.8in]{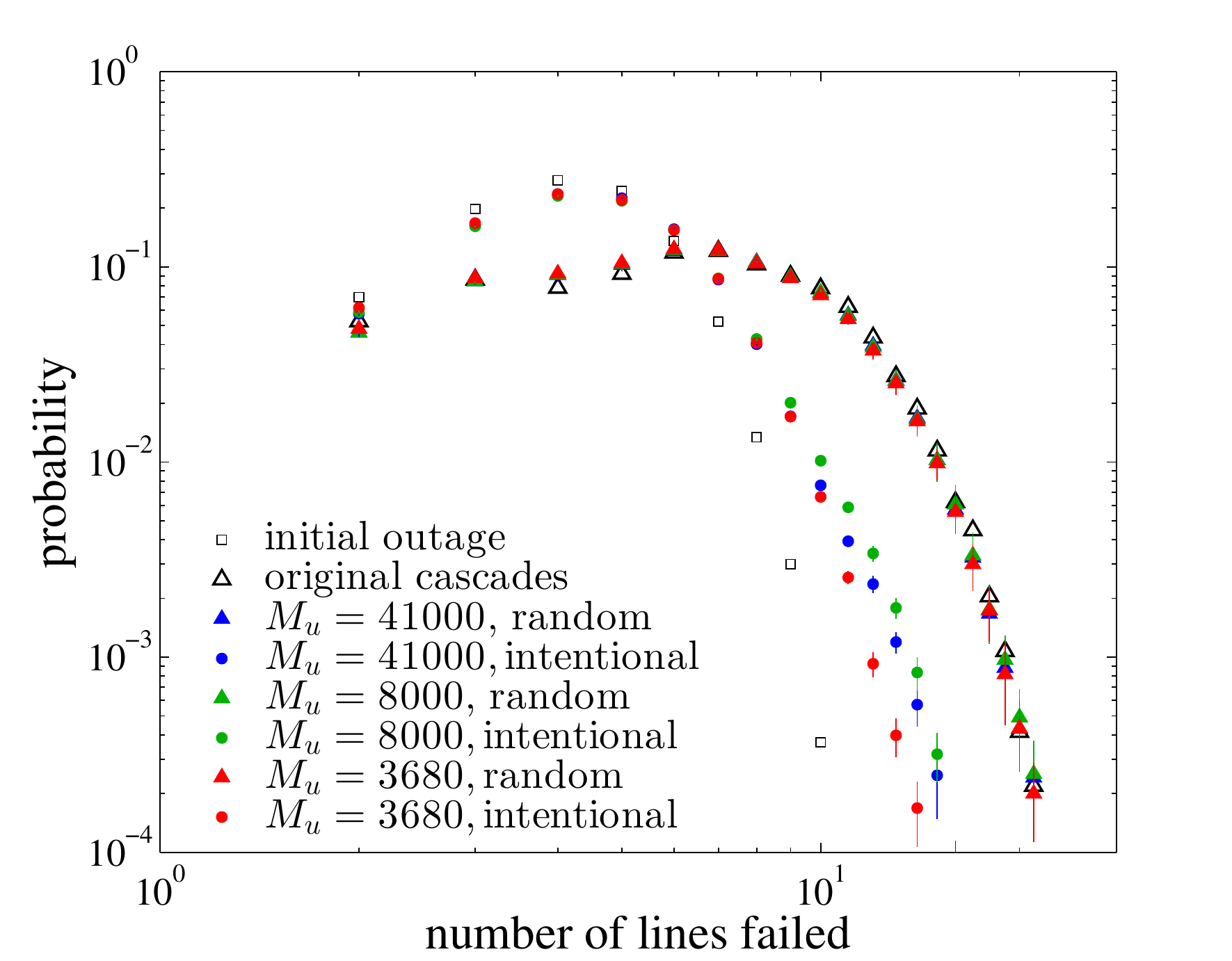}
\caption{Probability distributions of the total number of line outages
under two mitigation strategies. Triangles indicate total numbers of line outages of 
the original cascades under no mitigation; blue, green, and red triangles and dots
separately indicate total numbers of line outages of the simulated cascades under random and intentional mitigation for using 41000, 8000, and 3680 original cascades to quantify interactions; vertical lines indicate the standard deviations for 20 times of interaction model simulations; squares indicate initial line outages. }
\label{mitigate}
\end{figure}

Fig. \ref{mitigate} shows the probability distributions of total line outages 
under the two mitigation strategies for $M_u=41000$, $M_u=8000$, and $M_u=3680$.
It is seen that the risk of large-scale cascading failures 
can be significantly mitigated by weakening only a small number of key links. 
By contrast, the mitigation effect is minor if the same number of randomly chosen links are weakened.

This can be explained by the power law distribution of the link weights.
Most links have small weights and only a small number of links have much greater weights.
When randomly weakening links it is more possible to choose small-weight links 
and thus the total weights of the weakened links for random mitigation
can be significantly weaker than that for the intentional mitigation.
The probability for exactly choosing the key links and thus getting the largest possible total link weights for $M_u=41000$, $M_u=800$, and $M_u=3680$ are respectively as low as $3.10\times 10^{-29}$, $1.91\times 10^{-24}$, and $5.93\times 10^{-24}$.
The summation of the weights of weakened links for intentional and random mitigation are 
listed in Table \ref{mitigateSummary}, in which $m_{I_{\mathcal{L}^{\textrm{key}}}}$ 
and $\sigma_{I_{\mathcal{L}^{\textrm{key}}}}$ separately denote the average value and standard deviation of the total weights of the weakened links.
We can see that the total weights of the weakened links for intentional mitigation is more than a order greater than those for random mitigation. The standard deviations of the total weights of the weakened links for random mitigation are big because the distribution of the link weights follows power law and the link weights can vary significantly. The big standard deviations cannot be decreased by doing a large number times of simulation, which indicates that the random mitigation strategy is not stable and the mitigation effects between different random mitigation can be quite different.

\begin{table}[H]
\renewcommand{\arraystretch}{1.3}
\captionsetup{labelsep=space,font={footnotesize,sc}}
\caption{\\Link Weights for Different Mitigation Strategies}
\label{mitigateSummary}
\centering
\begin{tabular}{cccccc}
\hline
model & mitigation strategy & $M_u$ & $m_{I_{\mathcal{L}^{\textrm{key}}}}$ & $\sigma_{I_{\mathcal{L}^{\textrm{key}}}}$ \\
\hline
interaction & intentional & 41000 & 121114 & 0 \\
interaction & random & 41000 & 6126 & 6624 \\
interaction & intentional & 8000 & 23380 & 0 \\
interaction & random & 8000 & 1737 & 1570 \\
interaction & intentional & 3680 & 11247 & 0 \\
interaction & random & 3680 & 1043 & 696 \\
\hline
\end{tabular}
\end{table}

To quantitatively compare the effects of different mitigation measures,
the branching process is applied to estimate the average propagation of
the original and simulated cascades under two mitigation strategies.
The average value and standard deviation of the estimated average propagation are listed in Table \ref{lambda1}.
It is seen that the average propagation decreases dramatically
under intentional mitigation while decreases only a little 
under random mitigation. Also the relative standard deviations of the average propagation for random
mitigation are much higher than those for intentional mitigation, which can be explained by the big standard deviation of of the total weights of the weakened links for random mitigation.

\begin{table}[H]
\renewcommand{\arraystretch}{1.3}
\captionsetup{labelsep=space,font={footnotesize,sc}}
\caption{\\Average Propagation for Different Mitigation Strategies}
\label{lambda1}
\centering
\begin{tabular}{cccccc}
\hline
model & mitigation strategy & $M_u$ & $M$ & $m_{\hat{\lambda}}$ & $\sigma_{\hat{\lambda}}$ \\
\hline
interaction & intentional & 41000 & 41000 & 0.0966 & 0.000657 \\
interaction & random & 41000 & 41000 & 0.391 & 0.0128 \\
interaction & intentional & 8000 & 41000 & 0.113 & 0.00110 \\
interaction & random & 8000 & 41000 & 0.394 & 0.0154 \\
interaction & intentional & 3680 & 41000 & 0.0965 & 0.000711 \\
interaction & random & 3680 & 41000 & 0.391 & 0.0143 \\
\hline
\end{tabular}
\end{table}

In order to simulate the implementation of the mitigation strategy 
by weakening some links in real systems we add a relay blocking module in AC OPA model.
For intentional or random mitigation, when the source vertices of 
the predetermined links fail and the destination vertices of corresponding links
become overloaded and will be tripped by protective relays,
the probability of the operation of the relays will be reduced by 90\% and AC OPA will have a much greater chance to go to the next inner iteration without tripping this overloaded destination vertex line, in which AC OPF is performed to simulate the re-dispatching of generation and some loads are shed if necessary in order to eliminate the violation of the line limits.
In this way the AC OPA simulations can get cascades under mitigation strategies.

\begin{figure}[!t]
\captionsetup{justification=raggedright,singlelinecheck=false}
\centering
\includegraphics[width=2.8in]{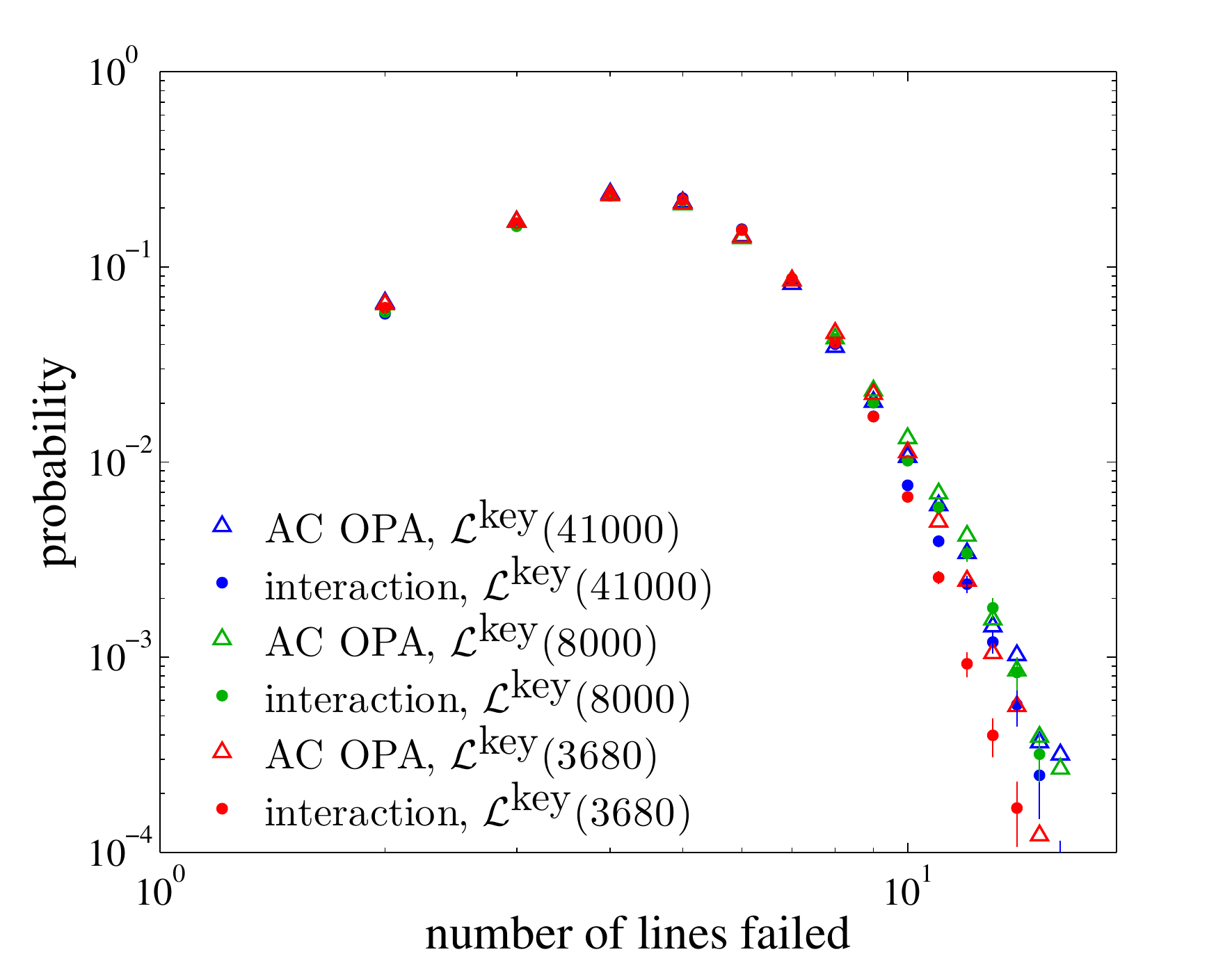}
\caption{Probability distributions of the total number of line outages under intentional mitigation for AC OPA and interaction model simulations. Blue, green, and red triangles indicate total numbers of line outages for AC OPA model simulation; blue, green, and red dots indicate total number of line outages for simulated cascades using 41000, 8000, and 3680 original cascades to quantify interactions; vertical lines indicate the standard deviations for 20 times of interaction model simulations. }
\label{mitigate1}
\end{figure}

In Fig. \ref{mitigate1} we compare the average probability distributions of the total numbers of the line outages 
for AC OPA model and interaction model under intentional mitigation for $M_u=41000$, $M_u=8000$, and $M_u=3680$.
Note that the identified key links for $M_u=41000$, $M_u=8000$, and $M_u=3680$ are separately denoted by $\mathcal{L}^{\textrm{key}}(41000)$, $\mathcal{L}^{\textrm{key}}(8000)$, and $\mathcal{L}^{\textrm{key}}(3680)$ and the AC OPA model simulation is performed for three times, each of which respectively weakens $\mathcal{L}^{\textrm{key}}(41000)$, $\mathcal{L}^{\textrm{key}}(8000)$, and $\mathcal{L}^{\textrm{key}}(3680)$.
It is seen that for all three cases using different $M_u$ the distributions for both models match very well under the intentional strategies.
Results for random mitigation are similar and thus are not given.

The branching process is also applied to estimate the average propagation of
the AC OPA and interaction model under intentional mitigation strategy, 
which are listed in Table \ref{lambda2}.
The average propagations for two models match each other very well and 
the standard deviations $\sigma_{\hat{\lambda}}$ of the average propagation are very small, indicating that 
first quantify interaction by using fewer original cascades (such as 8000 or 3680 cascades) and then performing the interaction model simulation to get 41000 cascades can get consistent and almost the same results as directly simulating 41000 cascades with the AC OPA model.

\begin{table}[!t]
\renewcommand{\arraystretch}{1.3}
\captionsetup{labelsep=space,font={footnotesize,sc}}
\caption{\\Average Propagation for AC OPA and Interaction Model}
\label{lambda2}
\centering
\begin{tabular}{cccccc}
\hline
model & mitigation strategy & $M_u$ & $M$ & $m_{\hat{\lambda}}$ & $\sigma_{\hat{\lambda}}$ \\
\hline
\scriptsize AC OPA  & \tabincell{c}{intentional \\$(\mathcal{L}^{\textrm{key}}(41000))$} & -- & 41000 & 0.0944 & -- \\
interaction & \tabincell{c}{intentional \\ $(\mathcal{L}^{\textrm{key}}(41000))$} & 41000 & 41000 & 0.0966 & 0.000657 \\
\scriptsize AC OPA  & \tabincell{c}{intentional \\ $(\mathcal{L}^{\textrm{key}}(8000))$} & -- & 41000 & 0.109 & -- \\
interaction & \tabincell{c}{intentional \\ $(\mathcal{L}^{\textrm{key}}(8000))$} & 8000 & 41000 & 0.113 & 0.00110 \\
\scriptsize AC OPA  & \tabincell{c}{intentional \\ $(\mathcal{L}^{\textrm{key}}(3680))$} & -- & 41000 & 0.0998 & -- \\
interaction & \tabincell{c}{intentional\\ $(\mathcal{L}^{\textrm{key}}(3680))$} & 3680 & 41000 & 0.0965 & 0.000711 \\
\hline
\end{tabular}
\end{table}

\subsection{Efficiency}  \label{efficiency}

In this section the improvement of efficiency brought about by the proposed interaction model is discussed. The timing for simulating 41000 cascades for AC OPA simulation and the interaction model simulation 
based on different number of original cascades $M_u$ is shown in Table \ref{index2}, in which $T_1$, $T$, and $T_2$ are respectively the time for AC OPA simulation, for calculating the probability that each component fails in generation 0 and the interaction matrix, and interaction model simulation. 
We can see that it is more time efficient to first quantify the interactions between the component failures 
with $M_u \ll M$ original cascades and then perform the interaction model simulation than it is to directly simulate $M$ cascades with AC OPA model.

\begin{table}[!t]
\renewcommand{\arraystretch}{1.3}
\captionsetup{labelsep=space,font={footnotesize,sc}}
\caption{\\Efficiency Improvement of Interaction Model}
\label{index2}
\centering
\begin{tabular}{cccccc}
\hline
model & $M_u$ & $M$ & $T_1$(hour) & $T$(second) & $T_2$(second) \\
\hline
\scriptsize AC OPA & -- & 41000 & 57 & 0 & 0  \\
interaction & 41000 & 41000 & 57 & 95  & 29 \\
interaction & 8000 & 41000 & 11 & 15  & 29 \\
interaction & 3680 & 41000 & 5 & 7  & 29  \\
\hline
\end{tabular}
\end{table}

When many times of simulation need to be performed the advantage of first quantifying the interactions and then
doing the interaction model simulation will become more obvious.
Assume it takes $t_1$ and $t_2$ to generate one cascade from AC OPA model and the interaction model and we have $t_1 \gg t_2$. 
As in Table \ref{index2}, the time for calculating the probability that each component fails in generation 0 and the interaction matrix by using $M_u$ cascades is denoted by $T(M_u)$.
To get $N$ sets of $M$ cascades, the ratio between the simulation time for AC OPA model and that 
for first quantifying the interactions and then performing interaction model simulation is
\begin{equation}
R=\frac{N M t_1}{M_u t_1 + T(M_u)+N M t_2}
\end{equation} 
and by letting $N M \rightarrow \infty$ we can get
\begin{equation}
\lim\limits_{N M \rightarrow \infty} R=\lim \limits_{N M \rightarrow \infty} \frac{t_1}{\frac{M_u t_1}{N M}+\frac{T(M_u)}{N M} + t_2}=\frac{t_1}{t_2}
\end{equation} 
which indicate a significant efficiency improvement for the interaction model simulations compared with the AC OPA simulations.

\begin{figure}[!t]
\captionsetup{justification=raggedright,singlelinecheck=false}
\centering
\includegraphics[width=2.8in]{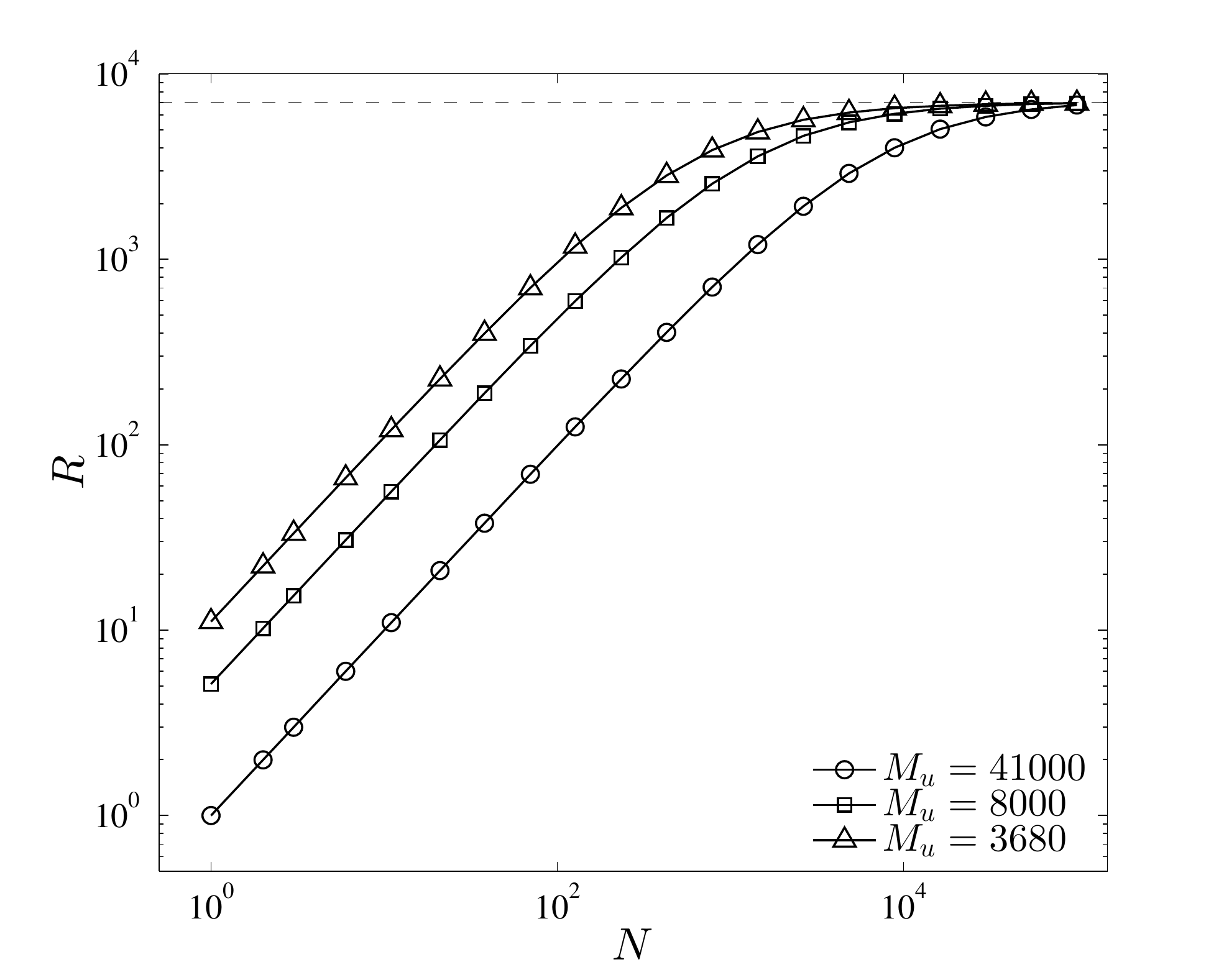}
\caption{$R$ for different $N$. Circles, squares, and triangles indicate the efficiency improvement $R$ for using 41000, 8000, and 3680 original cascades. The dash horizontal line indicates $R=7200$. }
\label{efficiencyplot}
\end{figure}

In our case $t_1 \simeq 5.04s$ and $t_2 \simeq 0.0007s$. Thus $\lim_{N M\rightarrow \infty}R \simeq 7200$. 
By letting $M=M^{\textrm{min}}=41000$, we show how $R$ changes with $N$ for $M_u=41000$, $M_u=8000$, and $M_u=3680$ in Fig. \ref{efficiencyplot}, in which we can see that $R$ first quickly increases and then finally saturates at about 7200. Also as expected, when $N$ is not large enough the efficiency improvement for $M_u=3680$ is better than that for $M_u=8000$ because smaller $M_u$ will lead to shorter time for obtaining original cascades from AC OPA simulation and also shorter time for quantifying the interactions.

The efficiency improvement can be reflected in studying the effects of mitigation measures. 
The interaction model can generate cascades and study the influence of component interactions on cascading failure risk much more time efficiently while reserving most of the general properties of the cascades.
In section \ref{mitigationResult} we have shown that the simulated cascades from interaction model
are consistent with the AC OPA simulation under the same mitigation strategy.
However, in order to obtain the cascades under a specific mitigation strategy
AC OPA simulation will need about 57 hours to obtain  41000 cascades 
while the interaction model simulation only requires about 29 seconds to obtain the same number of cascades
by simply changing some elements in the already obtained interaction matrix.
This efficiency improvement is important if we would like to quickly find out the impact of a mitigation measure.

Further, as mentioned in section \ref{model}, the proposed interaction model can be used for online decision-making support by fast predicting the consequences of the events happening in the system.
For online operation there is no enough time for performing detailed cascading failure simulations.
But we can obtain the interaction matrix offline from simulations of detailed cascading failure models or 
from statistical utility data and then apply the interaction model to quickly find out
the components or areas of the system that will most probably be affected so that a fast response can be performed
to pull the system back to normal conditions and to avoid or at least reduce the economic and social losses.
The efficiency improvement of the interaction model reflected in this aspect is a great advantage compared with 
many other cascading failure models.

\section{Conclusion} \label{conclusion}

In this paper we quantify the interaction between component failures
by calculating the probability that one component failure causes another 
and obtain the interaction matrix and interaction network.
Key links and key components are identified and an interaction model is proposed to simulate cascading failures 
and study how interactions between component failures influence cascading failure risk.

The interaction quantifying method and interaction model 
are validated to be able to capture general properties of the original cascades.
It is much more time efficient to first quantify the interactions between the component failures 
with fewer original cascades from more detailed cascading failure model, such as AC OPA, and then perform the interaction model simulation than it is to directly simulate a large number of cascades with a more detailed model.

The obtained interaction network can reveal the non-local property of the propagation of
cascading failures, which has been seen in several blackouts in real power systems. 
An obvious power law is found in the distributions of the link weights 
and the vertex out-strength and in-strength of the interaction network, 
suggesting that a small number of links and components are much more crucial than the others.
Cascading failure risks can be greatly mitigated by weakening a few key links, which can be implemented in real systems by wide area protection that blocks the operation of relays of the lines corresponding to 
the destination vertices of key links when the lines corresponding to the source vertices are tripped.

Further, the proposed interaction model can also be used for online decision-making support 
by predicting the consequences of the events based on the interaction matrix
obtained offline from simulations of more detailed cascading failure model or 
from statistical utility data.




\section*{Acknowledgment}

We gratefully thank anonymous reviewers for their insightful advice 
that helped greatly improve our paper.


%

\begin{IEEEbiography} [{\includegraphics[width=1in,height=1.25in,clip,keepaspectratio]{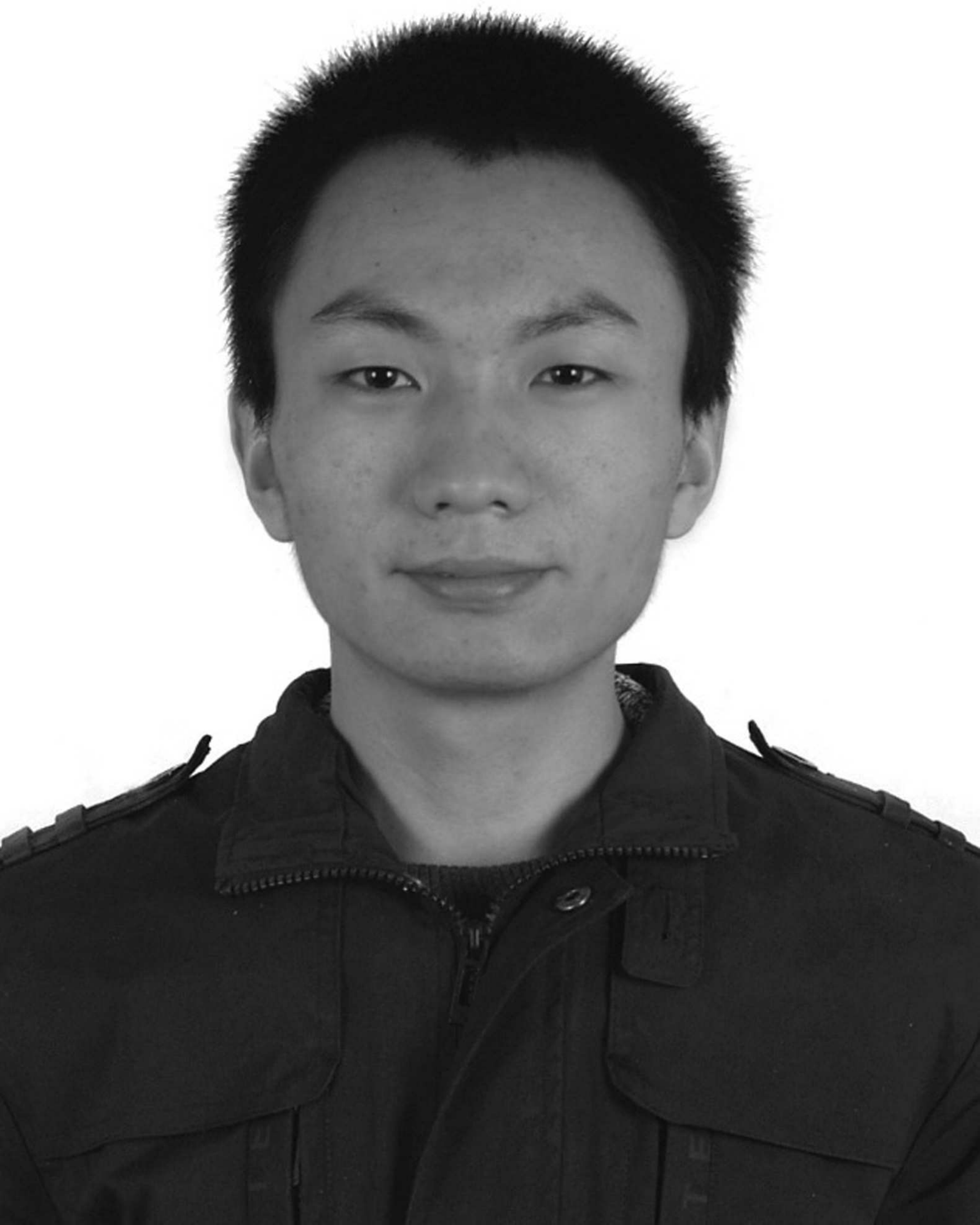}\vfill}]
{Junjian Qi} (S'12--M'13)
received the B.E. and Ph.D. degree both in Electrical Engineering from Shandong University in 2008 and Tsinghua University in 2013. He visited Prof. Ian Dobson's group at Iowa State University in Feb.--Aug. 2012 and is currently a research associate at Department of EECS, University of Tennessee in Knoxville.

His research interests include blackouts, cascading failure, state estimation, and synchrophasors.
\end{IEEEbiography}

\begin{IEEEbiography} [{\includegraphics[width=1in,height=1.25in,clip,keepaspectratio]{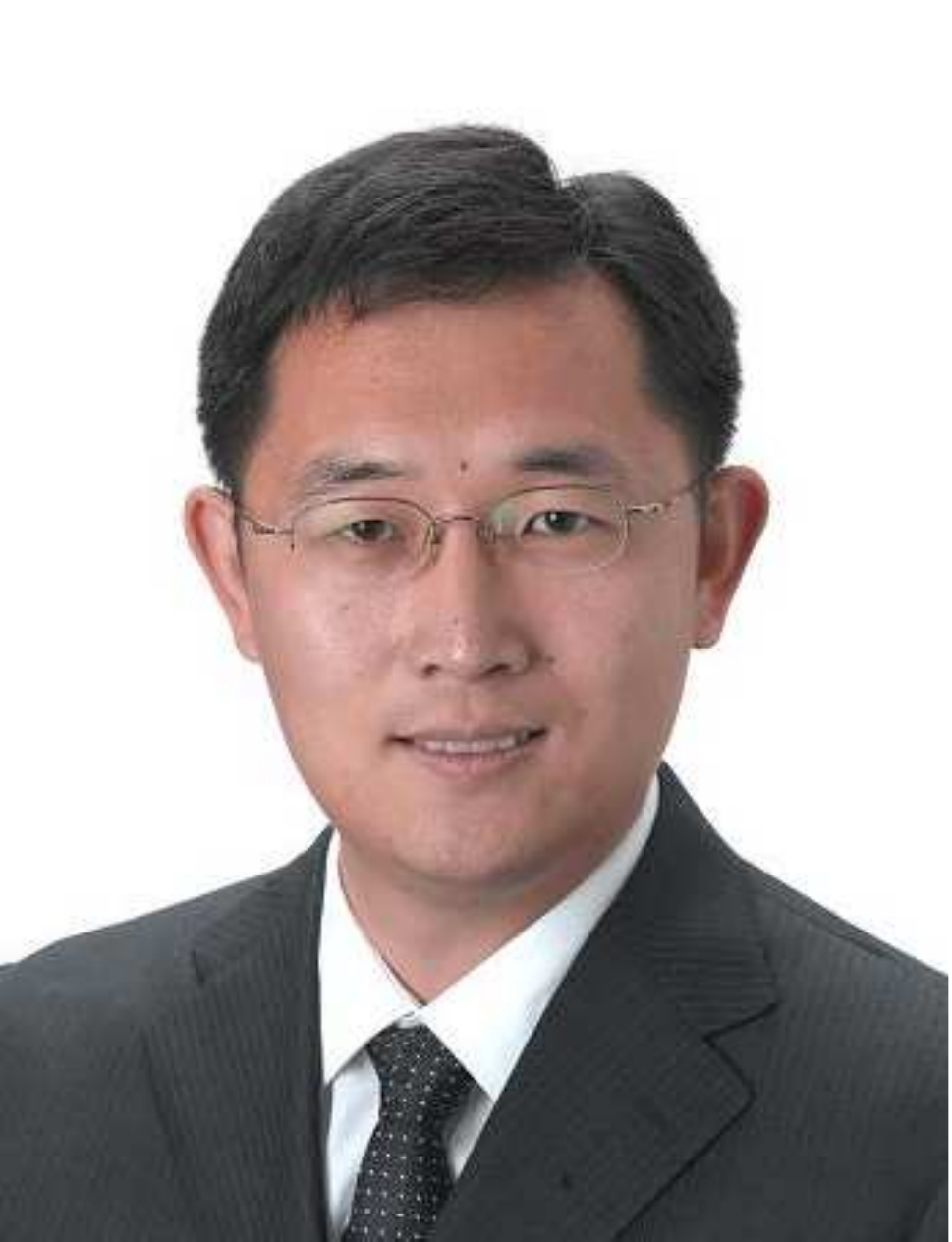}\vfill}]
{Kai Sun} (M'06--SM'13)
received the B.S. degree in automation in 1999 and the Ph.D. degree in control 
science and engineering in 2004  both  from Tsinghua University, Beijing, China.  
He was a postdoctoral research associate at Arizona State University, Tempe, from 2005 to 2007, 
and was a project  manager in grid operations and planning areas at EPRI, Palo Alto, CA
from 2007 to 2012. 

He is currently an assistant professor at the Department of EECS, University of Tennessee in Knoxville.
\end{IEEEbiography}

\begin{IEEEbiography} [{\includegraphics[width=1in,height=1.25in,clip,keepaspectratio]{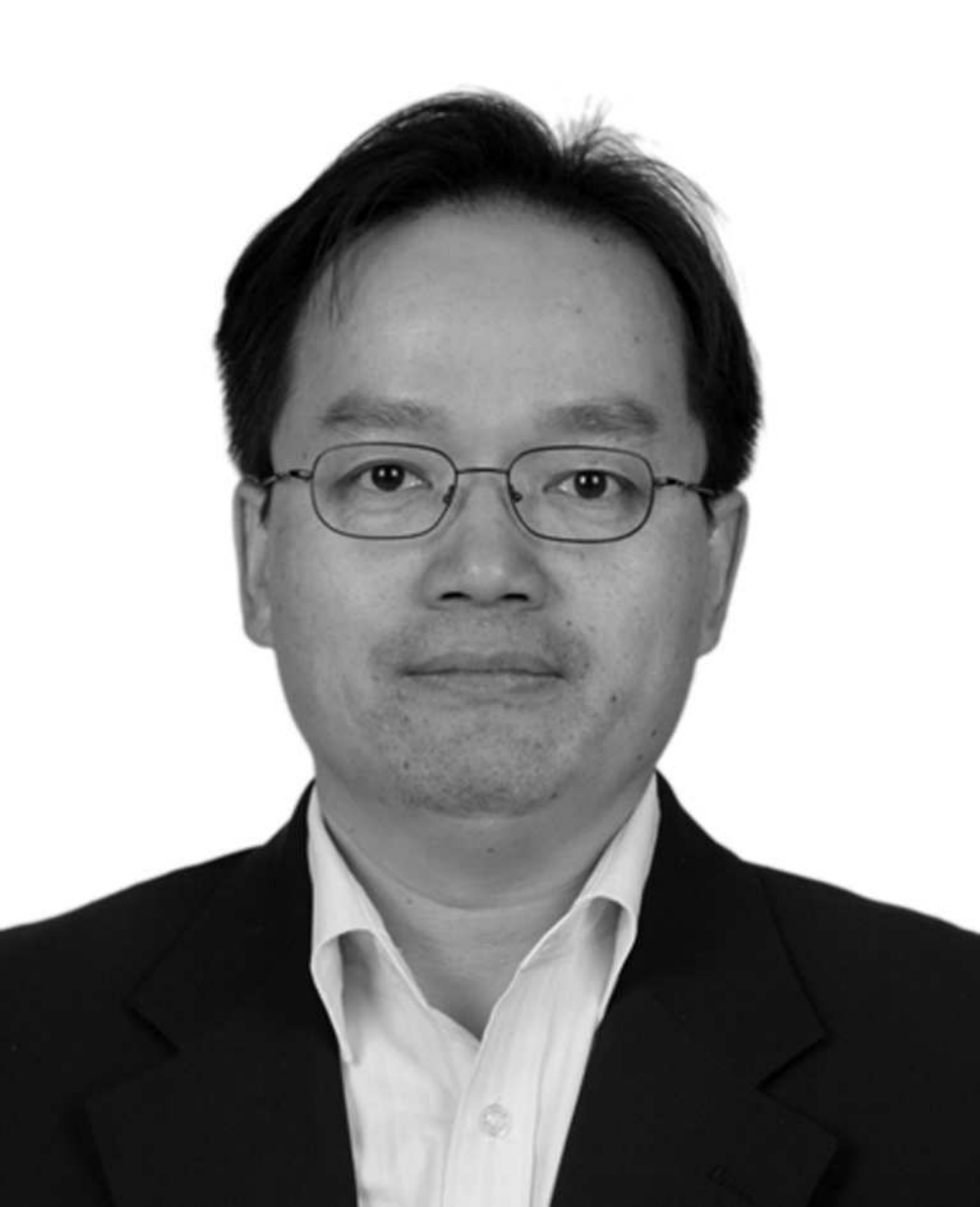}\vfill}]
{Shengwei Mei} (SM'05)
received the B.S. degree in mathematics from Xinjiang University, the M.S. degree in operations research from Tsinghua University, and the Ph.D degree in automatic control from the Chinese Academy of Sciences, in 1984, 1989, and 1996, respectively. 

He is currently a professor at Tsinghua University.
His research interests include power system analysis and control.
\end{IEEEbiography}

\end{document}